\documentclass[prb,floatfix,showpacs,superscriptaddress,longbibliography,twocolumn]{revtex4-2}

\usepackage[english]{babel}
\usepackage{amsmath,amssymb,amsfonts}
\usepackage{epsfig}
\usepackage{graphicx}
\usepackage{outlines}
\usepackage[dvipsnames]{xcolor}

\usepackage[colorlinks=true,linkcolor=blue,citecolor=red]{hyperref}
\usepackage[normalem]{ulem}

\DeclareMathAlphabet\mathbfcal{OMS}{cmsy}{b}{n}

\newcommand{\equ}[1]
{Eq.~(\ref{#1})}

\newcommand{\figu}[1]
{Fig.~\ref{#1}}

\newcount\colveccount
\newcommand*\colvec[1]{
  \global\colveccount#1
  \begin{pmatrix}
    \colvecnext
  }
  \def\colvecnext#1{
    #1
    \global\advance\colveccount-1
    \ifnum\colveccount>0
    \\
    \expandafter\colvecnext
    \else
  \end{pmatrix}
  \fi
}

\newtoks\rowvectoks
\newcommand{\rowvec}[2]{%
  \rowvectoks={#2}\count255=#1\relax
  \advance\count255 by -1
  \rowvecnexta}
\newcommand{\rowvecnexta}{%
  \ifnum\count255>0
  \expandafter\rowvecnextb
  \else
  \begin{pmatrix}\the\rowvectoks\end{pmatrix}
  \fi}
\newcommand\rowvecnextb[1]{%
  \rowvectoks=\expandafter{\the\rowvectoks&#1}%
  \advance\count255 by -1
  \rowvecnexta
}

\def\bcen{\begin{center}}
\def\ecen{\end{center}}

\def\a{\alpha}       \def\b{\beta}

\def\fig1{\hbox{\msytw Q}}

\def\=={\equiv}

\def\qed{\raise1pt\hbox{\vrule height5pt width5pt depth0pt}}
\def\iome{i\omega_n}

\def\cG0{{\cal G}_0}
\def\cG{{\cal G}}

\def\up{\uparrow}  \def\dw{\downarrow}
 
\def\ka{{\bf k}}

  \def\Im{\mbox{Im}}
 
 \def\=={\equiv}
 
\def\Im{{\rm Im}} \def\Re{{\rm Re}} 
 \def\ep0{\epsilon_{p}} \def\ed0{\epsilon_{d}}

\def\ka{{\bf k}}

\usepackage{bbold}
\def\11{\mathbb{1}}
\def\00{\mathbf{0}}

\begin{document}
\title{Local vs non-local correlation effects in interacting quantum spin Hall insulators} 
\author{L.~Crippa}\email [] {lorenzo.crippa@physik.uni-wuerzburg.de}
\affiliation{Institut f\"ur Theoretische Physik und Astrophysik and W\"urzburg-Dresden Cluster of Excellence ct.qmat, Universit\"at W\"urzburg, 97074 W\"urzburg, Germany}
\affiliation{Scuola Internazionale Superiore di Studi Avanzati (SISSA),
Via Bonomea 265, 34136 Trieste, Italy}

\author{A.~Amaricci}\email [] {amaricci@sissa.it}
\affiliation{CNR-IOM DEMOCRITOS, Istituto Officina dei Materiali,
Consiglio Nazionale delle Ricerche,
Via Bonomea 265, 34136 Trieste, Italy}
\affiliation{Scuola Internazionale Superiore di Studi Avanzati (SISSA),
Via Bonomea 265, 34136 Trieste, Italy}

\author{S.~Adler}
\affiliation{Institut f\"ur Theoretische Physik und Astrophysik and W\"urzburg-Dresden Cluster of Excellence ct.qmat, Universit\"at W\"urzburg, 97074 W\"urzburg, Germany}
\affiliation{Institute of Solid State Physics, TU Wien, A-1040 Vienna, Austria}

\author{G.~Sangiovanni}
\affiliation{Institut f\"ur Theoretische Physik und Astrophysik and W\"urzburg-Dresden Cluster of Excellence ct.qmat, Universit\"at W\"urzburg, 97074 W\"urzburg, Germany}

\author{M. Capone}\email [] {capone@sissa.it}
\affiliation{Scuola Internazionale Superiore di Studi Avanzati (SISSA), Via Bonomea 265, 34136 Trieste, Italy}
\affiliation{CNR-IOM DEMOCRITOS, Istituto Officina dei Materiali,
Consiglio Nazionale delle Ricerche,
Via Bonomea 265, 34136 Trieste, Italy}

% \author{A.~Amaricci}
% \affiliation{CNR-IOM DEMOCRITOS, Istituto Officina dei Materiali,
% Consiglio Nazionale delle Ricerche, Via Bonomea 265, I-34136 Trieste, Italy}

% \author{M. Capone}
% \affiliation{Scuola Internazionale Superiore di Studi Avanzati (SISSA),
% Via Bonomea 265, 34136
% Trieste, Italy}

%\date{ \today }

\begin{abstract}
The impact of Coulomb interaction on the electronic properties of a
quantum spin-Hall insulator is studied using quantum cluster methods,
disentangling local from non-local effects.
We identify different regimes, according to the value of the
bare mass term, characterized by drastically
different self-energy contributions.
% We identify regions of parameters characterized by drastically
% different self-energy contributions, according to the value of the
% bare mass term.
For small mass, non-local correlations start to be important and
eventually dominate over local ones when getting close enough to the
zero-mass semi-metallic line. For intermediate and large mass, local
correlation effects outweigh non-local corrections, leading to a
first-order topological phase transition, in agreement with
previous predictions.
\end{abstract}

\maketitle

The description of solids is based on many-electron wave functions
built out of Bloch states, i.e. of delocalized eigensolutions of
single electrons in a periodic potential~\cite{Bloch1929}.
The resulting electronic energy-momentum dispersion is determined by
the quantum mechanical amplitude of electrons hopping from site to
site of the lattice. Another important ``atomistic'' ingredient
influencing the bandstructure is the relative energetic alignment of
the local levels in the material.
Electronic wave functions with strong contribution from atoms whose
outer shell is lower in energy form mainly valence bands, while the
conduction bands are mostly associated to orbitals of the less
electronegative elements.
If a clearcut distinction in the above sense exists for a set of bands
of an insulator at all momenta in the Brillouin zone, then the
bandstructure can be adiabatically connected to some atomic limit and
is therefore classified as topologically trivial.
%\note{I think this short paragraph about tensile HgTe drifts away from the logic of the introduction.}

Non-trivial topological effects can, on the contrary, arise when there is a 
sizeable entanglement between conduction and valence electron wave functions 
in momentum space. 
This is the situation in the well-known case of tensile strained 
HgTe~\cite{Dai2008,Brune2011}. In this II-VI semiconductor, the 6$s$ level of
the Hg is empty and forms to a large extent the conduction band. Yet, due
to its large atomic number, Hg is subject to sizeable
relativistic corrections, that at the $\Gamma$ point ``pull'' some of
its 6$s$ character down to the valence
eigenvalues~\cite{BHZ2006,Rothe2010}.
This gap inversion is distinct from any atomic limit and the resulting bandstructure
can hence be classified by a non-zero global topological
$\mathbb{Z}_2$-invariant~\cite{Kane2005PRLa,Kane2005PRL,FuKane2007,Kane2008NP,Hasan2010RMP}

An electron-electron interaction brings about many-body physics and leads to a breakdown of the independent-particle picture~\cite{Abrikosov1959,Hubbard1,Hubbard2,Hubbard3,Mott1949}.
In many cases a bandstructure can still be identified, notwithstanding visible lifetime effects that broaden the single-particle eigenvalues of an amount proportional to the electron-electron scattering rate.
This mechanism is encoded in the electronic self-energy which, in addition to the broadening, describes $i$) a renormalization of the energy positions of the local level and $ii$) a modification of the bare hopping amplitudes.

The first of these two effects, namely the correlation-induced
modulation of the local atomic splitting ($i$), can induce a change of
the topological invariant, a direct repercussion of the aforementioned
connection between band ordering and atomic level energies. This has
been analyzed in detail within dynamical mean-field theory (DMFT)~\cite{Kotliar2004PT,GeorgesKotliar1996} and
non-local extensions thereof~\cite{Capone2004,Park2008,Sakai2012}, as well as with quantum variational
approaches~\cite{Potthoff2003PRL,Aichhorn2006,Senechal2008,Balzer2009,Seki2018}.
Within the so-called Bernevig-Hughes-Zhang-Hubbard (BHZH)~\cite{BHZ2006}
model, the trivial-to-nontrivial transition driven by the competition
between the bare local splitting $M$ separating the two orbitals and
the Hubbard interaction strength $U$ has been extensively
studied~\cite{Hohenadler2013ROPIP,miyakoshiPRB87,Wang2012EEL,Yoshida2012PRB,Hohenadler2011,Amaricci2017PRB,Budich2012PRB,Amaricci2018PRB,Rachel2018ROPIP}.

The many-body effects on the electron hopping processes, i.e. $ii$,
have been instead investigated much less, so far. Since it is crucial
to describe the interplay between $i$ and $ii$ on an equal footing, in
this work we use a cluster extension of DMFT~\cite{Capone2004,Park2008,Sakai2012} to quantify
the non-local self-energy renormalizations of the bare hopping amplitudes in the BHZH model, for the case of an average occupation of two electrons per site.
Our conclusions, summarized in Sec. ~\ref{sec:phasediag}, are to some extent unexpected, considering the fact that we are focusing on the two-dimensional model: 
the region of the phase diagram where local many-body corrections of the type $i$ dominate is fairly extended, in particular for large local orbital energy splitting $M$ and large interaction $U$. In addition, we unveil the existence of a regime where non-local corrections prevail and small-${\bf q}$ instabilities can be expected. Interestingly, such strongly non-local solutions are found for small values of $M$, i.e. deep inside the topologically non-trivial phase. Our results thus point towards the existence of a quantum spin Hall insulator in which an onsite repulsion influences primarily the non-local physics.

The paper is organized as follows: after introducing the model in
Sec.~\ref{Sec:Model}, we discuss in Sec.~\ref{Sec:NonLocalAndHtop} the
corresponding self-energy structure obtained within cluster dynamical
mean-field theory (CDMFT). In Sec.~\ref{sec:resultsA} we present
results for parameters close to the semimetallic line whereas in
Sec.~\ref{Sec:TQPT} we focus on the transition from the trivial band-
to the non-trivial quantum spin Hall insulator driven by the mass term
at intermediate-sizeable values of the Coulomb repulsion. In
Sec.~\ref{sec:phasediag} we assemble our results in a phase diagram
and highlight the relative importance between local and non-local
self-energy corrections. Before the conclusions, we elaborate on the 
comparison between the BHZH model and the Kane-Mele-Hubbard
(KMH) one. 

\section{Model and methods}\label{Sec:Model}
\begin{figure}[ht]
  \includegraphics[width=\linewidth]{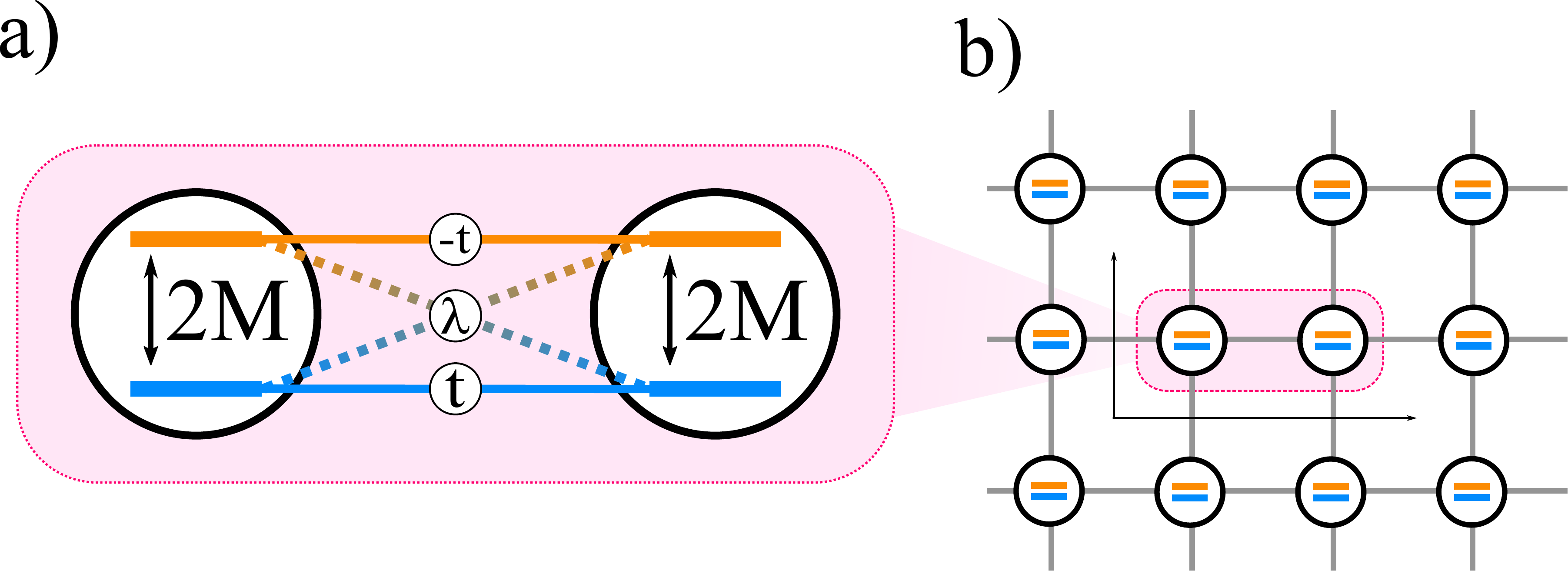}
  \caption{(Color online)
    (a) Schematic representation of the Hamiltonian of the two-site cluster we use in our CDMFT calculations.  (b) Embedding of the cluster in the two-dimensional lattice. 
%    ffective impurity model for the CDMFT using ``replica'' bath levels. Each bath level $i$ is formed by a copy of the impurity cluster and parametrized by $M_i$, $t_i$ and $\lambda_i$. The bath Hamiltonian has the same structure of the local cluster one. Each bath replica is coupled to the impurity via an amplitude $V_i$. (c) Reference system for the VCA method. The cluster local Hamiltonian is treated variationally to find stationary points of the self-energy functional. The variational parameter is the coupling $V$ to the auxiliary bath levels.
  }  
  \label{fig1}
\end{figure}

We consider an interacting BHZ model~\cite{BHZ2006} in two dimensions.
Using two-component spinors 
$\Psi(\mathbf{k})=(
\mathrm{c}_{\ka a\sigma}\,,
\mathrm{c}_{\ka b\sigma})$ where $\mathrm{c}_{\ka a\sigma}$ is the
annihilation operator for an electron with lattice momentum ${\bf k}$
in orbital $m=a,b$ with spin $\sigma=\uparrow,\downarrow$
we can  write the single-particle part of the Hamiltonian as 
\begin{equation}
\sum_{\ka}\Psi_{\ka \uparrow}^\dag {H}(\ka)\Psi_{\ka \uparrow} + \sum_{\ka}\Psi_{\ka \downarrow}^\dag {H}^*(-\ka) \Psi_{\ka \downarrow}
\end{equation}
  where
\begin{equation}
\begin{split}
  {H}(\ka)=
  &\lambda \sin k_x \tau_x+\lambda \sin k_y \tau_y+\\
  &\big[M-2t(\cos k_x +\cos k_y)\big]\tau_z
  \label{2d-bhz-hk}
\end{split}
\end{equation}
is written in terms of Pauli matrices in the orbital subspace
$\vec\tau$. The relation between the ${H}(\ka)$ of the two blocks with
opposite spins follows from time-reversal symmetry (TRS).
The model describes two orbitals with local energies respectively at
$\pm M$.  $t$ is a standard nearest-neighbor orbital-diagonal hopping
while $\lambda$ is a nearest-neighbor orbital-off-diagonal spin-orbit
coupling (SOC) -- See \figu{fig1}.
We fix the total occupation per site to 2 (half-filling), set
$\epsilon\equiv 2t$ as our energy unit and choose $\lambda=0.3$.

We assume a local interaction within the two-orbital manifold. In
particular we use the density-density version of the Slater-Kanamori
interaction~\cite{Kanamori1963,Georges2013ACMP}, analogous to previous works on the
same model. This interaction depends on a Hubbard repulsion $U$ and a
Hund's coupling~\cite{Medici2011PRB,Georges2013ACMP} $J$ describing the interaction between electrons on the same orbital with
opposite spins and on different orbitals with same spin,
respectively. In our calculations we further consider $J=U/4$.

%\begin{equation}
%\begin{split}
%  {H}_\mathrm{int}=&U\sum_{il}n_{im\uparrow}n_{im\downarrow}+\\
%&(U-2J)\sum_{ia\neq m^{'}}n_{im\uparrow}n_{im^{'}\downarrow}+\\
%&(U-3J)\sum_{im< m^{'}}(n_{im\uparrow}n_{im^{'}\uparrow}+n_{im\downarrow}n_{im^{'}\downarrow}).
%\end{split}
%\label{hint-kanamori-bhz-2d}
%\end{equation}
%where $U$ and $J$ parameterize the interaction and $n_{im\sigma}=c^+_{im\sigma}c_{im\sigma}$ is the density operator on site $i$ for the orbital $m$ and spin $\sigma$. These interaction form corresponds to the density-density part of the two-orbital Kanamori Hamiltonian ~\cite{kanamori}. In the rest of this work we will set $J=U/4$.

In the absence of interaction, the half-filled model is characterized
by a topological phase transition separating a quantum spin-Hall
insulator for $M < 2 \epsilon$ and a trivial band insulator for $M>2
\epsilon$.
%In order to account for the interaction, a simple analytical solution is not feasible and it is necessary to resort to approximations or numerical methods.
%In presence of interaction the general solution of the model can only be obtained using numerical methods.
Previous analyses have used single-site dynamical mean-field theory
(DMFT)~\cite{GeorgesKotliar1996,metzvol,MullerHartmann1989} to include interaction effects
non-perturbatively~\cite{miyakoshiPRB87,Wang2012EEL,Yoshida2012PRB,Budich2012PRB,Amaricci2015PRL,Amaricci2017PRB,Amaricci2018PRB}.
Within DMFT the self-energy is purely local, while retaining the full
frequency dependence. The main effect of the interactions is hence a
renormalization of the local level splitting, described by the mass
term $M$. %encoded in the zero-frequency value of the self-energy.
Further, it was previously found that the frequency dependence of the
DMFT self-energy influences the theremodynamics of the topological
phase transition, which turns from continuous to discontinuous upon
increasing the value of the bare Coulomb repulsion $U$~\cite{Amaricci2015PRL,Amaricci2016PRB,Weyl2020PRR}.

Here, by means of methods which include short-ranged dynamical
correlations~\cite{Maier2005}, namely cluster dynamical mean-field
theory~\cite{Kotliar2001,Biroli2002} (CDMFT) and variational cluster
approximation (VCA)~\cite{Potthoff2003,Potthoff2003PRL} we go beyond
DMFT and assess the parameters' region in which the effect of the
non-local self-energy corrections are strongest.

\section{Cluster-DMFT, the self-energy and  the topological Hamiltonian}\label{Sec:NonLocalAndHtop}

In CDMFT~\cite{Kotliar2001,Biroli2002}, a real-space extension of DMFT, the lattice model is mapped onto an interacting cluster embedded in a self-consistent bath.
Here, we solve the cluster impurity model by means of exact
diagonalization~\cite{amaricci2021edipack}.
This requires an approximation of the bath in terms of a finite and small number of energy levels~\cite{Senechal2010}. 
In order to enforce the lattice symmetries, one can organize the bath
in a handful of ``replicas'' of the impurity cluster, each sharing the
same symmetry~\cite{Capone2004,CivelliThesis,Koch2008}.
A schematic representation of the extended impurity problem is
presented in \figu{fig1}. In this work we limit to the
smallest two-site cluster and we consider a cluster aligned along the
$x$ direction. This obviously introduces an artificial difference
between \textit{x} and \textit{y}, but it reduces the computational cost, allowing for a
reasonable size of the bath (two replicas) and for a systematic
investigation of the model parameters.

%MC improve
%We will show in the following the results for the \textit{x} cluster, though the ones for the \textit{y} one are completely analogous. This cluster choice, of course, is not ideal in that it breaks the original symmetry of the square lattice. However, since -as we will discuss- the self-energy shows the same symmetries as the Hamiltonian, it is rather safe to assume that its effects would be uniform in the two directions, if the more suitable square plaquette was chosen.

\begin{figure*}
  \includegraphics[width=\linewidth]{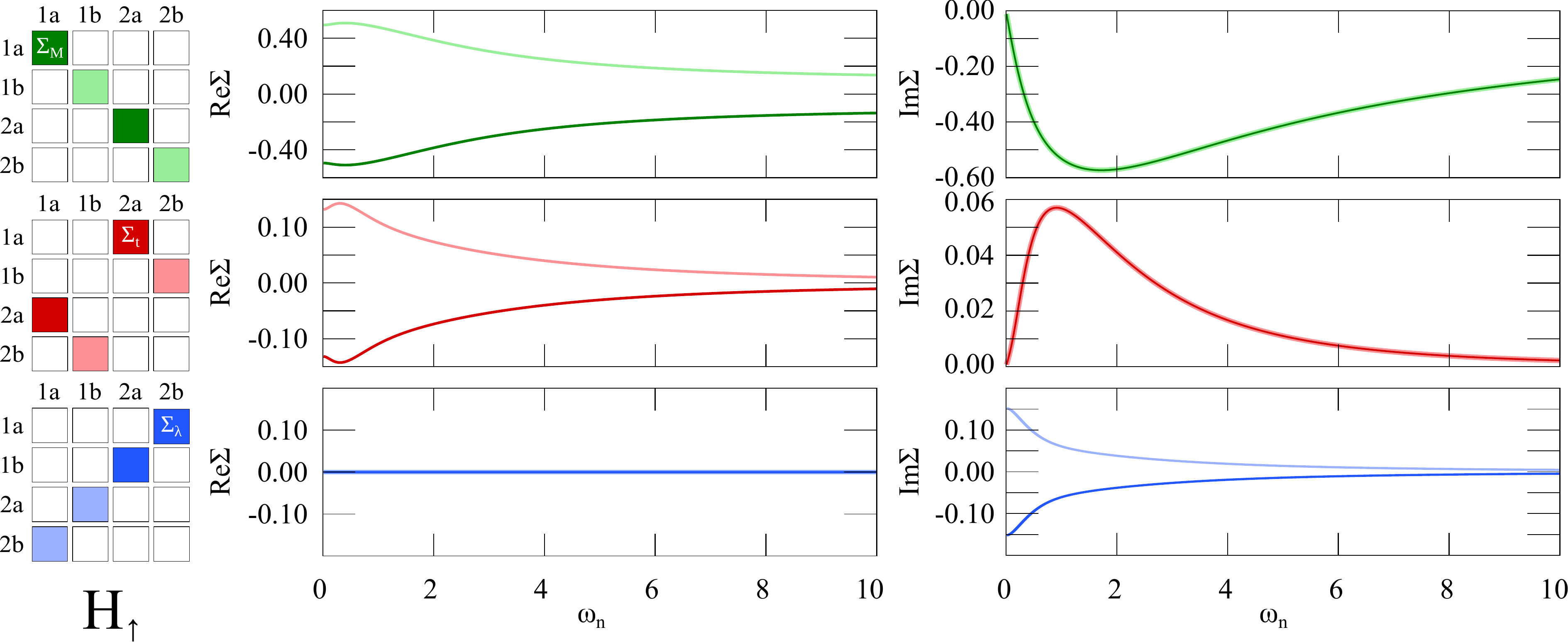}
  \caption{ (Color online)  Behavior of the cluster self-energy $\mathbf{\Sigma}_{i\a\up,j\b,\up}(\iome)$ on the Matsubara axis for the spin-$\up$ block. The components are indicated in the left column with the colors denoting their symmetries. The cluster sites are labeled by $i,j=1,2$ and the orbitals by roman letters $\a,\b=a,b$. The results shown are from CDMFT calculations performed at $U=3$ and $M=0.8$, inside the QSHI phase. The spin-$\dw$ block's behavior is dictated by symmetries, while all the remaining components vanish within any periodization scheme. Figure adapted from~\cite{CrippaThesis}.}
  \label{fig2}
\end{figure*}

Within CDMFT we directly obtain the cluster self-energies and Green's
function. Yet, since the CDMFT cluster has open boundary conditions,
in order to build lattice observables it is necessary to use a
periodization procedure. In this work we employ a  $\Sigma$-scheme periodization~\cite{Kotliar2001,Capone2004} where the lattice self-energy (omitting the orbital index) is obtained as
%reproducing periodically the cluster self-energy according to
\begin{equation}
\begin{split}
\Sigma(\mathbf{k},z)=\dfrac{1}{N_{C}}\sum_{ij}^{N_{C}}\mathrm{e}^{-i\mathbf{k}\cdot (\mathbf{r}_{i}-\mathbf{r}_{j})}[\mathbf{\Sigma}]_{ij}(z).
\end{split}
\label{periodization}
\end{equation}
with  $[\mathbf{\Sigma}]_{ij}$ the self-energy submatrix connecting sites $i$ and $j$ of the cluster. 
This choice of periodization is particularly suited for phases in
which the self-energy is regular, as it happens for the topological
and trivial insulating states. Other periodizations based on the
Green's function and the cumulant
~\cite{Senechal2000,Stanescu2006,Sakai2012} are more suited for the
study of the Mott insulator and its proximity. We have verified that
in the relevant region of parameters that we describe below, the three
periodizations yield qualitatively similar results.

The  self-energy function $\mathbf{\Sigma}$ in CDMFT has a
2$\times$2 (for orbitals) times $N_C$$\times N_C$ matrix structure, for each
spin. For our $N_C = 2$ cluster, we have therefore a 4$\times$4 matrix
with orbital and site indices, as depicted in the left panels of
Fig. \ref{fig2}. In these panels we emphasize that the
non-zero components of the cluster self-energy mirror the symmetry of
the single-particle Hamiltonian and only three independent functions
of the frequency are needed to describe the full self-energy. We label
these three functions as $\Sigma_M(\iome)$, $\Sigma_t(\iome)$ and
$\Sigma_ {\lambda}(\iome)$ since they appear as additive corrections
to the respective non-interacting parameters. 
In the $\Sigma$-periodization, the three independent functions coincide with the real space components of the cluster self-energy shown in Fig. \ref{fig2}. For example, the middle
panels of the figure show
Re$\Sigma_t(\iome)$=Re$\Sigma_{1a,2a}$=Re$\Sigma_{2a,1a}$=$-$Re$\Sigma_{1b,2b}$=$-$Re$\Sigma_{2b,1b}$
while the corresponding imaginary parts are identical. Notice that all
imaginary parts other than $\Sigma_\lambda(\iome)$ vanish in the zero-frequency limit.

We can then express the periodized lattice self-energy in terms of the three
independent lattice components introduced above: 
\begin{equation}
  \begin{split}
    {\Sigma}(\mathbf{k},\iome)=&
    \Re{\Sigma}_M(\iome)\tau_z + i\Im{\Sigma}_M(\iome)\tau_0 + \\
    &\left[\Re{\Sigma}_t(\iome)\tau_z +
      i\Im{\Sigma}_t(\iome)\tau_0\right]\cos k_{x}-\\    
    &\Im{\Sigma}_{\lambda}(\iome)\tau_x\sin k_{x}
  \end{split}
\label{sigma-periodized}
\end{equation}

Comparing with the non-interacting Hamiltonian, we see that $\Sigma_M(\iome)$, $\Sigma_t(\iome)$ and $\Sigma_{\lambda}(\iome)$ play the role of dynamical renormalizations of the three parameters $M$, $t$ and $\lambda$ respectively. 
The form of eq.\eqref{sigma-periodized} naturally emerges in the $\Sigma$ periodization, but it is obtained also with different periodizations as it directly stems from symmetry arguments. 
%However, we note that the structure of the self-energy, in its original cluster form, mimics the symmetries of the Hamiltonian expressed in the same base. It then follows that the dependence of the self-energy on momentum is analogous in all periodization schemes, with the difference resting in the values of the frequency-dependent complex coefficients that in the $\Sigma$-scheme coincide with $\Sigma_{M,t,\lambda}(i\omega_{n})$. In particular, within the Fermi liquid regime (everywhere but in the Mott region) the imaginary part of the diagonal component of the self-energy vanishes for $\omega\rightarrow0$.

To diagnose the topological properties of interacting systems, various schemes based on effective Hamiltonians as well as on Green's functions~\cite{Hofstetter2018,Markov2019,thunstrom2019topology} have been formulated. 
Here, we make use of the well-established \textit{topological Hamiltonian}~\cite{WangZhang2012}, defined as
\begin{equation}
  {H}_\mathrm{topo}(\mathbf{k})={H}(\mathbf{k})+{\Sigma}(\mathbf{k},\omega=0).
  \label{Htop}
\end{equation}

The effect of the interaction on the topological properties is hence described by the zero-frequency limit of ${\Sigma}(\ka,\omega)$. Since inversion symmetry is preserved, the construction of the topological index can be further simplified~\cite{FuKane2007,Wang2012}, by evaluating the parity of the eigenstates of the occupied bands at the four time-reversal invariant momenta of the square BHZ Brillouin zone. 

% The $\ZZZ_2$ topological invariant $\nu$ then results from the expression:
%\begin{equation}
%(-1)^\nu = \prod_{i,\alpha}\sqrt{p_{i\alpha} }
%\end{equation}
%where we have chosen $\sqrt{-1}=+i$ and defined the indices $i$ and $a$ to run over the TRIMs ($[0/\pi,0/\pi]$) and occupied bands respectively.
%\lc{An alternative way of determining the interacting topological invariant is the so-called pole expanded Hamiltonian~\cite{karsten}, which takes into account the topological character of the auxiliary bath levels. This method, which acquires interest in our case due to the replica structure of the bath, constitutes a relevant expansion direction for this work.}

\section{Local and non-local correlations and the competition between $M$ and $U$} \label{sec:results}

%The Hubbard-Bernevig-Hughes-Zhang is characterized by five model parameters, $t$, $\lambda$, $M$,  $U$ amd $J$. In the non-interacting limit, the topological transition is controlled essentially by the ratio $M/t$, even if the parameter $\lambda$ is necessary fot the gap protection. The Hubbard $U$ and its ratio with the hopping scale $t$ measure the strength of the interaction which leads to electron-electron correlations. The Hund's coupling $J$ is important to shape the correlation-driven part of the phase diagram, as discussed in previous work~\cite{} and we fix it to a value such that a Mott insulating solution is actually realized.

The physics of the BHZH model is mostly determined by the interplay and competition between $U$ and $M$. The avoided-crossing for $M<2\epsilon$ leading to the band gap is guaranteed by the orbital off-diagonal non-local hybridization $\lambda$, whose precise value does not however affect the overall phase diagram. The ratio $J/U$ determines the critical interaction for the occurrence of the Mott transition but it does not qualitatively influence the properties of the quantum spin Hall phase obtained at intermediate values of $U$ and $M$.
The limits of very large $M$ and $U$ are expected to host a band- and a Mott insulator, respectively, two topologically trivial phases.% which are well understood%
For this reason we focus here on the most interesting region where $U$ and $M$ are comparable and study the relevance as well as the effect of local and non-local correlations.
%From here on, the $\Sigma_{M,t,\lambda}$ coefficients will refer to $\omega=0$. To allow the eigenvalue to have the form \eqref{eigenvalues-with-sigma}, only the real parts of $\Sigma_{M}$ and $\Sigma_{t}$ and the imaginary part of $\Sigma_{\lambda}$ are nonzero at $\omega\rightarrow 0$. As a consequence of the chosen cluster, the bare parameters coupling to $k_{y}$ are not renormalized.
%The strength of the relative components of the self-energy will therefore potentially influence the deviation of the system from the noninteracting behaviour in every region of the $U-M$ phase diagram.\\
%Clearly, the level splitting, the hopping term and the spin-orbit coupling are now all affected by nonlocal correlations. 
In particular, we discuss two paradigmatic regions where we find remarkably different correlations effects. These are highlighted in red boxes in the phase diagram that concludes our analysis (\figu{fig7}): (A) the small $U$ and small $M$ parameter range, where a perturbative treatment of the interactions is possible, and (B) the large $U$ and $M$ parameter range, where we approach the topological transition and we have fingerprints of Mott-like strong correlations.

%In the large-U, large-M region, especially around the topological phase transition, the local component $\Sigma_{M}$ dominates, as shown by the green gradient in \figu{fig7}. The red gradient, instead, represents the relative strength of the nonlocal components $\Sigma_{t}$ and $\Sigma_{\lambda}$-(which are comparable in modulus) with respect to the local one. This grows for decreasing M and is largest in the low-$U$ region before the Mott transition.\\
%In the two following subsections we will separately address the features of the interacting model in these two relevant regions.

\subsection{Small-to-intermediate interaction strength}\label{sec:resultsA}

\begin{figure}[ht]
  \includegraphics[width=\linewidth]{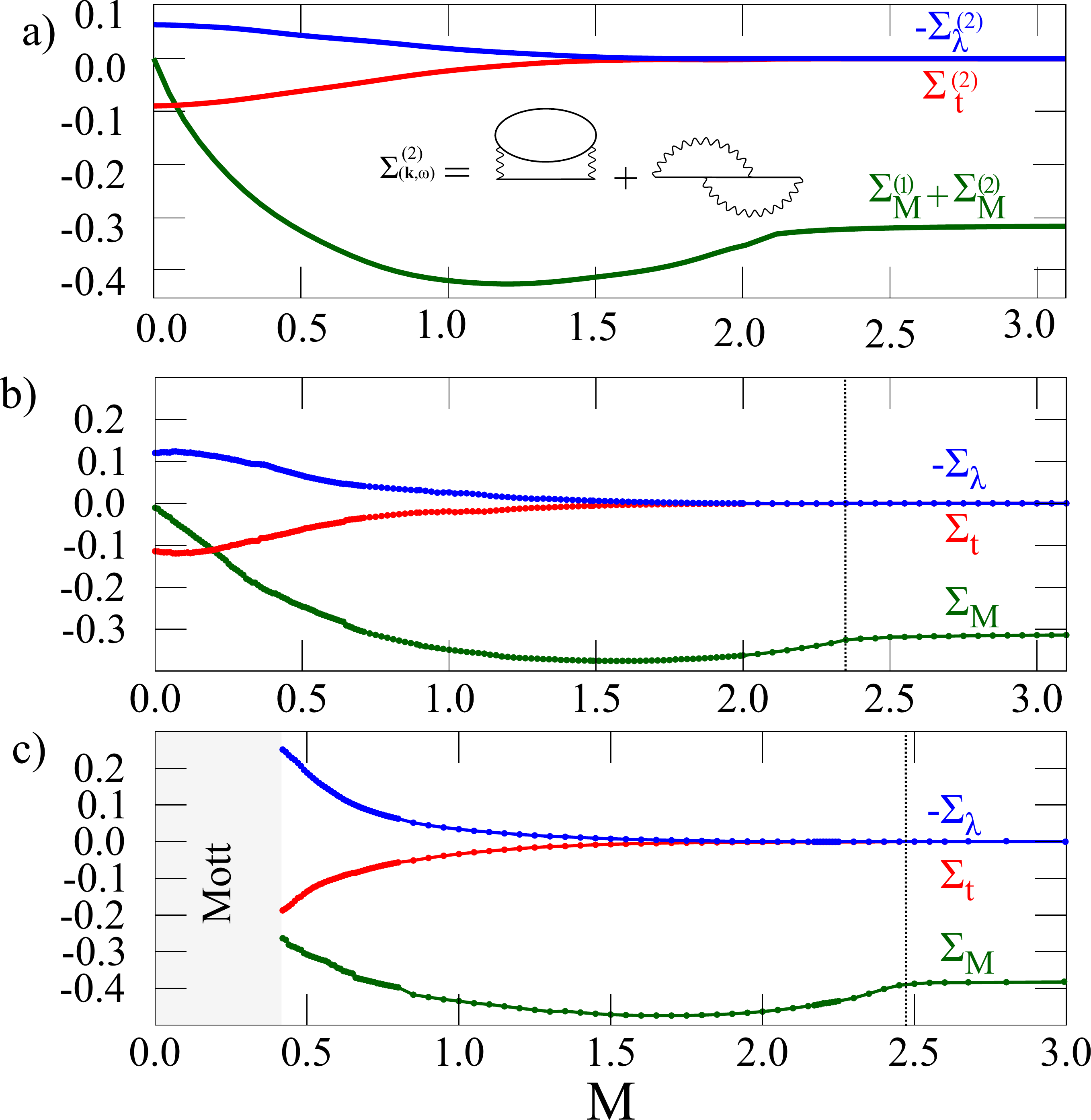}
  \caption{ (Color online) 
The zero-frequency limit of the three relevant components of $\Sigma$ as a function of $M$ for two different values of the interaction's strength: $U=2.5$ in panels \textbf{a)} and \textbf{b)} and $U=3.5$ in panel \textbf{c)}. For $U=2.5$, no Mott transition occurs while, for $U=3.5$, a Mott phase exists at small $M$ (see also \figu{fig7}). The topological phase transition driven by $M$ is indicated by the dotted vertical line. \textbf{a)} Self-energy from the lowest order diagrammatic expansion, as sketched by the Feynman diagrams. The system is gapless at the $[\pi,0]$ and $[0,\pi]$ high-symmetry points for $M=0$, and the local component $\Sigma_{M}$ vanishes, while $\Sigma_{t}$ and $\Sigma_{\lambda}$ remain finite. The CDMFT self-energy is shown in \textbf{b)} and \textbf{c)}. For both values of $U$, the main self-energy component at the topological phase transition is $\Sigma_{M}$ whereas $\Sigma_{t}$ and $\Sigma_{\lambda}$ start to dominate upon reducing $M$. Figure adapter from~\cite{CrippaThesis}.
  }
  \label{fig3}
\end{figure}

\begin{figure}[ht]
  \includegraphics[width=\linewidth]{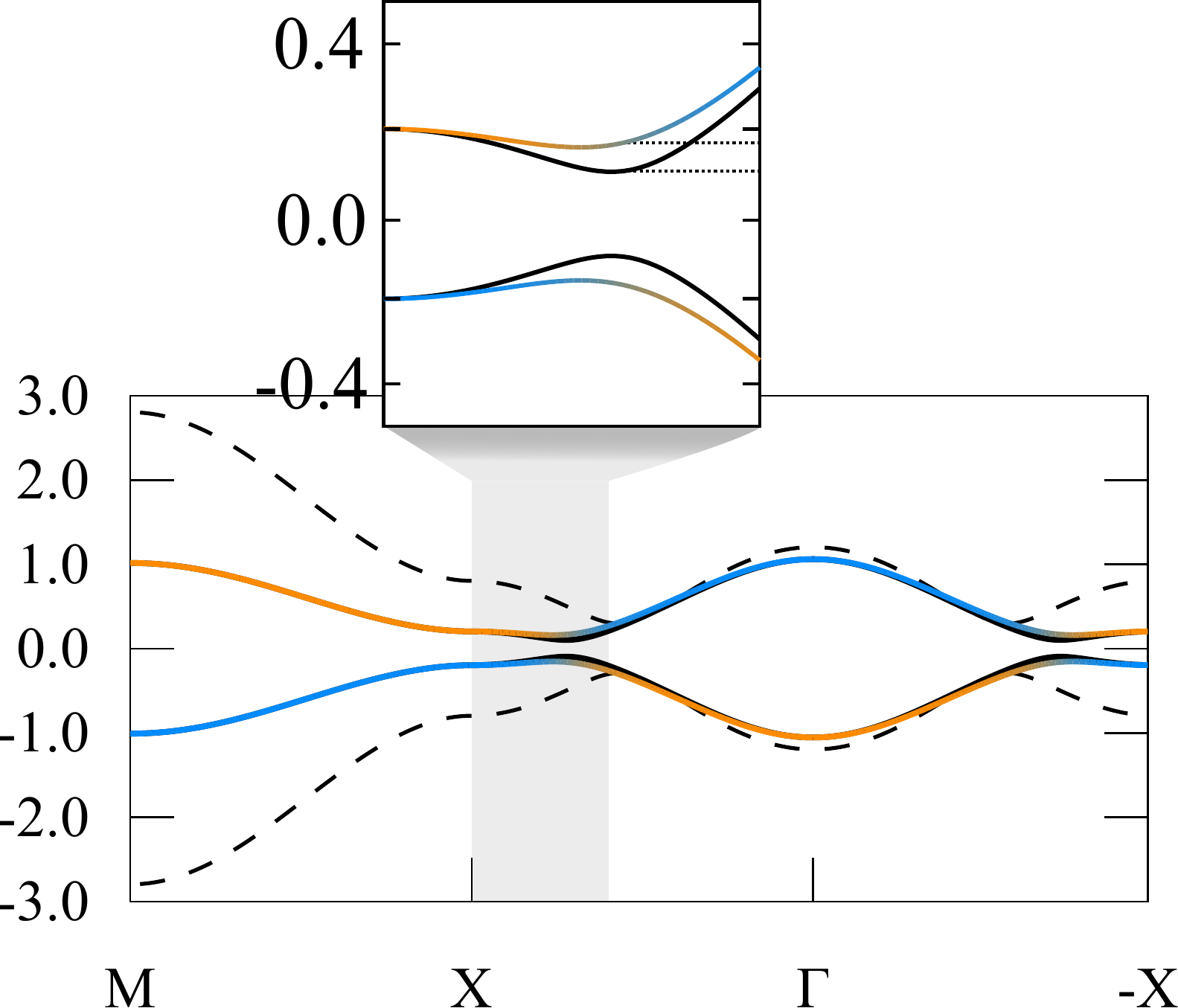}
  \caption{ (Color online) Effect of nonlocal correlations as described within CDMFT on the renormalized band structure of the model $Z_{\ka}H(\ka)$, in the QSHI region, for $U=3$ and $M=0.8$. The path in the Brillouin zone connects the high-symmetry points $M=[\pi,\pi]$, $\pm X=[\pm\pi,0]$ and $\Gamma=[0,0]$. The blue and orange bands show the renormalized dispersion in presence of all self-energy contributions, with the projection of the eigenvectors on the two orbitals denoted by the color intensities. The black solid line represents the band dispersion without the orbital off-diagonal component $\Sigma_{\lambda}$, while the dashed line is the bare non-interacting dispersion.
  }
  \label{fig4}
\end{figure}

%The low-U, low-M limit is deep in the QSHI phase. correlations do not modify the topology of the system in this region. They do, however, alter the single-particle excitation spectrum of the model; moreover, nonlocal effect acquire a much higher relevance here than in the previously outlined high-U high-M case. As it can be seen

In the spirit of the topological Hamiltonian formulation, in this
section and in the following we will study the effect of the
interactions through the zero-frequency values of the self-energies,
that renormalize the corresponding  Hamiltonian parameters. Using the
short-cut notation $\Sigma_l = \Sigma_l(\iome\to 0)$, with
$l=M,t,\lambda$, the eigenvalues of the topological Hamiltonian
\equ{Htop} reads~\footnote{The expression is manifestly non-symmetric
  due to the choice of the cluster which is aligned along the x
  direction, hence non-local self-energies enter only in the $k_x$
  direction. For a more symmetric cluster, the $k_x$ and $k_y$
  direction would display the same renormalization.}
\begin{equation}
  \begin{split}
    E^2_{\pm,\sigma}=&
    \left[(M+\Sigma_M)-(\epsilon-\Sigma_t)\cos k_{x} -\epsilon\cos k_{y}\right]^{2}+
    \\
    &(\lambda-\Sigma_\lambda)^{2}\sin^{2}k_{x}+\lambda^{2}\sin^{2}k_{y}.
\end{split}
\label{eigenvalues-with-sigma}
\end{equation}
In all the calculations presented in this work the three self-energies
are always negative on the positive Matsubara frequency axis. Hence, from Eq. (\ref{eigenvalues-with-sigma}),
we find that the effective mass is reduced by the interactions while
both non-local terms of the Hamiltonian, i.e. the SOC $\lambda$ and
the orbital-diagonal hopping $\epsilon$ are increased by the
self-energy corrections.

For small values of the interaction the self-energy function can be
determined by means of perturbation theory in the interaction U.
This allows to compare the CDMFT results to those obtained within a
controlled approximation retaining full momentum dependence. 
In particular, the contributions to the self-energy components $\Sigma_l$ can be
determined at each order at specific $k$ points.
% For small values of the interaction the three self-energy components
% can be obtained by means of perturbation theory in the
% interaction. 
The first-order term, i.e. the Hartree diagram, influences only
$\Sigma_M$.
The second-order diagrams contributing to the self-energy are depicted
in the inset to panel (a) of \figu{fig3}. These terms
contribute to all the self-energy components in
Eq.~\ref{eigenvalues-with-sigma}.  
In panels (a) and (b) of \figu{fig3} we compare indeed
the perturbative results for an intermediate value of $U = 2.5$  as a
function of $M$ with the CDMFT results for the same parameters. 
%Contrary to the cluster self-energies,% 
The perturbative self-energy $\Sigma^{(2)}(\mathbf{k},\omega)$ possesses full momentum dependence. Therefore, in order to compare it to the CDMFT results, we assume that \equ{sigma-periodized}, symmetrically extended in the $k_{y}$ direction, is a reasonable approximation for the $k$-dependent self-energy, and extract the $\Sigma_{M,t,\lambda}(i\omega)$ coefficients shown in \figu{fig3}a via a projection onto the corresponding lattice harmonics, i.e. by integrating over the Brillouin zone the quantities $\Sigma^{(2)}(\mathbf{k},\omega)$, $\Sigma^{(2)}(\mathbf{k},\omega)\cos k_x $ and $\Sigma^{(2)}(\mathbf{k},\omega)\sin k_x $ respectively.

%The self-energy components shown in panel (a) are extracted from the 
%full $\Sigma(\mathbf{k},\omega)$ at specific $k$-points, assuming a 
%momentum dependence of the type of Eq.~\ref{sigma-periodized}
%\footnote{More precisely, assuming the full momentum-dependent self-energy to be symmetric in the two $k$ directions, the components in \figu{fig3}(a) are evaluated at those $k$-points that make them comparable with the $\Sigma_{M,t,\lambda}$ coefficients in \eqref{sigma-periodized}: the off-diagonal $\lambda$ component is therefore evaluated at $\mathbf{k}=[\frac{\pi}{2},0]$; the diagonal momentum-independent component at $\mathbf{k}=[\frac{\pi}{2},\frac{\pi}{2}]$ and the diagonal momentum-dependent component is obtained as a difference between the previous value and the diagonal self-energy at $\mathbf{k}=[0,\frac{\pi}{2}]$.}

The agreement is surprisingly good, even at a quantitative level. 
We can thus infer that a perturbative expansion including local and non-local (momentum-dependent) diagrams is particularly accurate for a relatively large window of interaction.
It should be noted that we intentionally do not show a direct
comparison deep in the small-$U$ limit, as our numerical
implementation with a truncated bath is particularly well suited for
intermediate and large interactions and the small values of the
self-energies, inevitable for small $U$, challenge our numerical
accuracy.

%In \figu{fig3} the off-diagonal components drastically increase, while the diagonal one $\Sigma_{M}$ is reduced with decreasing M.\\ The noninteracting BHZ model at $M=0$ and half-filling is gapless at the $[\pi,0]$ and $[0,\pi]$ high-symmetry points. 

%From single-site DMFT studies on multi-orbital models ~\cite{parragh2013,giorgiojan2014} we know that $\Sigma_{M}$ is proportional to the bare $M$ parameter beyond the effects of the Hartree term, which is related to orbital polarization. Higher order, and especially second order, self-energy Feynman diagrams in U are crucial to determine the low-frequency structure of the $\Sigma_{\lambda}$ and $\Sigma_{t}$ components as well. A direct inspection of the $\Sigma^{(2)}$ terms (see sketch in gray square) is presented in the first panel of \figu{fig3}: here, the lowest order diagram contribution for each component have been plotted, them being respectively the Hartree and second order term for the $M$ component (minus a constant shift) and the second order terms alone for the nonlocal components. 

The physical picture is non trivial. For $M=0$ the non-local self-energies $\Sigma_t$ and $\Sigma_{\lambda}$ are finite while the local component $\Sigma_M$ vanishes~\cite{Parragh2013}. Increasing $M$ we find a rather rapid decrease of the non-local components, which vanish even before the topological transition is reached. On the other hand the absolute value of $\Sigma_M$  increases, as found in single-site DMFT, and reaches a maximum where $\Sigma_t$ and $\Sigma_{\lambda}$ vanish. For larger values of $M$, $\vert\Sigma_M\vert$ is slightly reduced and stays nearly constant in the trivial band insulator.

%As the behavior with respect to $M$ highlights, the corresponding self-energy component disappears for $M\rightarrow0$, while the other two saturate to constant values. We have checked that the nonlocal components are similarly reduced to 0 as the corresponding bare parameter vanishes.

%This behavior is retained in the full single-site~\cite{giorgiojan2014} and cluster DMFT picture, as shown in the second panel of \figu{fig3}: on the M=0 axis and before the MIT the interacting system preserves the semimetallic character, reflecting the symmetry between the orbitals. The green $\Sigma_{M}$ line tends to 0 for zero mass, while the $\Sigma_{t}$ and $\Sigma_{\lambda}$  components of the cluster self-energy remain finite. This latter effect is obviously beyond by the single-site picture, and constitute the biggest advantage in insight provided by cluster techniques.\\
If we further increase $U$ (\figu{fig3}c) the system is a Mott insulator at small $M$. Here, the self-energies diverge. Still, increasing $M$ we restore a QSH insulator (as already discuss in Refs.~\cite{Budich2013PRB,Amaricci2015PRL,Werner2007b}). There, the relative qualitative behavior between the various self-energy components discussed for smaller $U$ is recovered.

%, the evolution of the component undergoes a drastic change, all of them diverging at the transition. For small enough value of U, the behavior of the components before the transition still mimics the one outlined by the second order diagrams, with the $M$ component decreasing and the $t$ and $\lambda$ increasing.\\
%This scenario can be compared with that of \figu{fig5}: there, higher order diagrams are relevant in the overall shape of the self-energy, and the $M$ component no longer decreases upon reducing $M$, rapidly diverging to enter the Mott phase. 

As a result, we have a wide region of parameters where the QSHI is affected by local as well as non-local self-energies, beyond the single-site DMFT picture where only $\Sigma_M$ is nonzero. 
The bandstructure renormalized by all three CDMFT components, $\Sigma_t$, $\Sigma_{\lambda}$ and $\Sigma_M$ is plotted in \figu{fig4}. In particular, we show the eigenvalues of $Z_{\ka}H_{\mathrm{topo}}(\ka)$, where ${Z}_{\mathbf{k}}=\big[1-\tfrac{\partial{\Sigma}(\mathbf{k},\omega)}{\partial\omega}\big|_{\omega=0}\big]^{-1}$ is the quasiparticle weight measuring the coherence of the low-energy electronic states at different momenta.
%Hence This object takes into account the renormalization of the parameters, but also the lifetime of excitations through the quasiparticle weight $Z$, and is therefore suitable to describe the low energy single-particle excitations both in the QSHI and BI regimes, as show in 

%In \figu{fig4} we disentangle the effect of the different components of $\Sigma$.
%IMPROVE ----
% In each panel, the dotted line represents the band structure in presence of only some components of $\Sigma$, while the colored one is the fully renormalized dispersion.
%$\Sigma_{M}$ provides an uniform shrinking of the bandwidth of the system. The component $\Sigma_{t}$ opposes this behavior, enhancing the bandwidth as shown in region \textbf{a} of the first panel. The enhancement is, moreover, k-dependent since the $\Sigma_{t}$ component couples to $\cos k_{x}$.\\
%The topological bandgap is protected by virtue of $\lambda$. The large value of the corresponding component $\Sigma_{\lambda}$ has the effect of enhancing the bandgap, as shown in region \textbf{b}.\\

The comparison with the non-interacting results shows a non-uniform renormalization in momentum space. 
A direct effect of the momentum-dependence of the self-energy is visible close to the point where the inverted gap opens (as highlighted in the inset). There we clearly see an increase of the gap, which is easily traced back to the Re$\Sigma_{\lambda}$ component enhancing the effective spin-orbit coupling $\lambda_{eff}=\lambda-\Sigma_{\lambda}$ according to Eq. \ref{eigenvalues-with-sigma}.

%A note has to be made which highlights an unintended effect of the asymmetry of the chosen cluster: from the eigenvalue expression \eqref{eigenvalues-with-sigma}, it is apparent that the gap can indeed close at the $Y=[0,\pi]$ point if the if $M+\Sigma_{M}+\Sigma_{t}=0$. This is an unphysical result purely due to the asymmetric renormalization of the hopping amplitude in the two directions, that would disappear using a more symmetric cluster.

\subsection{Strong correlation regime}\label{Sec:TQPT} 

\begin{figure}[h]
  \includegraphics[width=\linewidth]{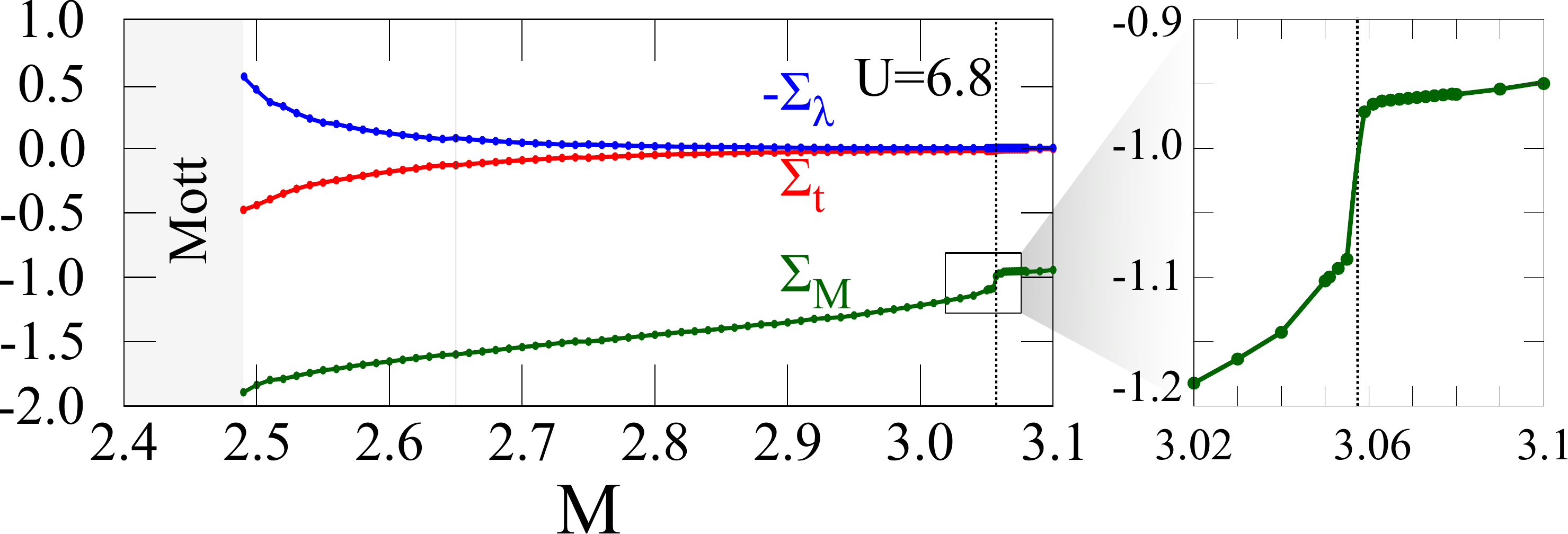}
  \caption{ (Color online) 
The three relevant components of $\Sigma$ as a function of $M$ for $U$=6.8. The topological phase transition occurs at the vertical dotted line. There, the only relevant self-energy component is $\Sigma_{M}$ and its first-order jump highlighted in the right panel.
  }
  \label{fig5}
\end{figure}

In this section we further increase $U$ and therefore stabilize the Mott phase. The Mott insulator is rather rigid against the increase of the local orbital splitting, hence we need a sizeable value of $M$ to stablize a QSHI (see \figu{fig5}). The qualitative behavior is quite similar to what we have shown in \figu{fig3}.
The transition from the Mott insulator to the QSHI upon increasing $M$ is however pushed to such a high value of $M$ that the small-$M$ region where $\Sigma_M$ would start to approach zero is completely hidden and we have significant local correlation effects in the whole QSHI phase (in the interval $2.48 \lesssim M \lesssim 3.06$ for the chosen value of $U$).

%Accordingly, it can be seen in \figu{fig5}, the renormalization effects along the transition line, and in particular for large U, are chiefly due to $\Sigma_{M}$. 

The CDMFT analysis reveals that, along the topological transition line and pretty irrespectively on $U$, the non-local parameters $t$ and $\lambda$ are almost unaffected by the interactions. Therefore the gap closing occurs at the $\Gamma$ point as soon as the first term in \eqref{eigenvalues-with-sigma} becomes zero, i.e. for
$
M+\Sigma_{M}=2\big(\epsilon-\frac{\Sigma_{t}}{2}\big).
$
This expression defines the topological transition line in \figu{fig7}. Due to the negligible value of $\Sigma_{t}$ in this specific range of parameters, we  recover the scenario revealed by single-site DMFT~\cite{Amaricci2015PRL}, where  the transition line  is given by $M+\Sigma_{M}=2\epsilon$ and it becomes first-order at a critical value of $U$ and $M$.

%Due to the local character of the renormalization, all the relevant effects presented by the single-site DMFT picture are retained in the cluster scenario: after a critical value of U ($\approx 6$), the renormalization provided by $\Sigma_{M}$ in the QSHI and BI regions becomes so different that the transition becomes discontinuous.\\

%\subsection{Energetics of the First-Order transition}

\begin{figure}%[h]
  \includegraphics[width=\linewidth]{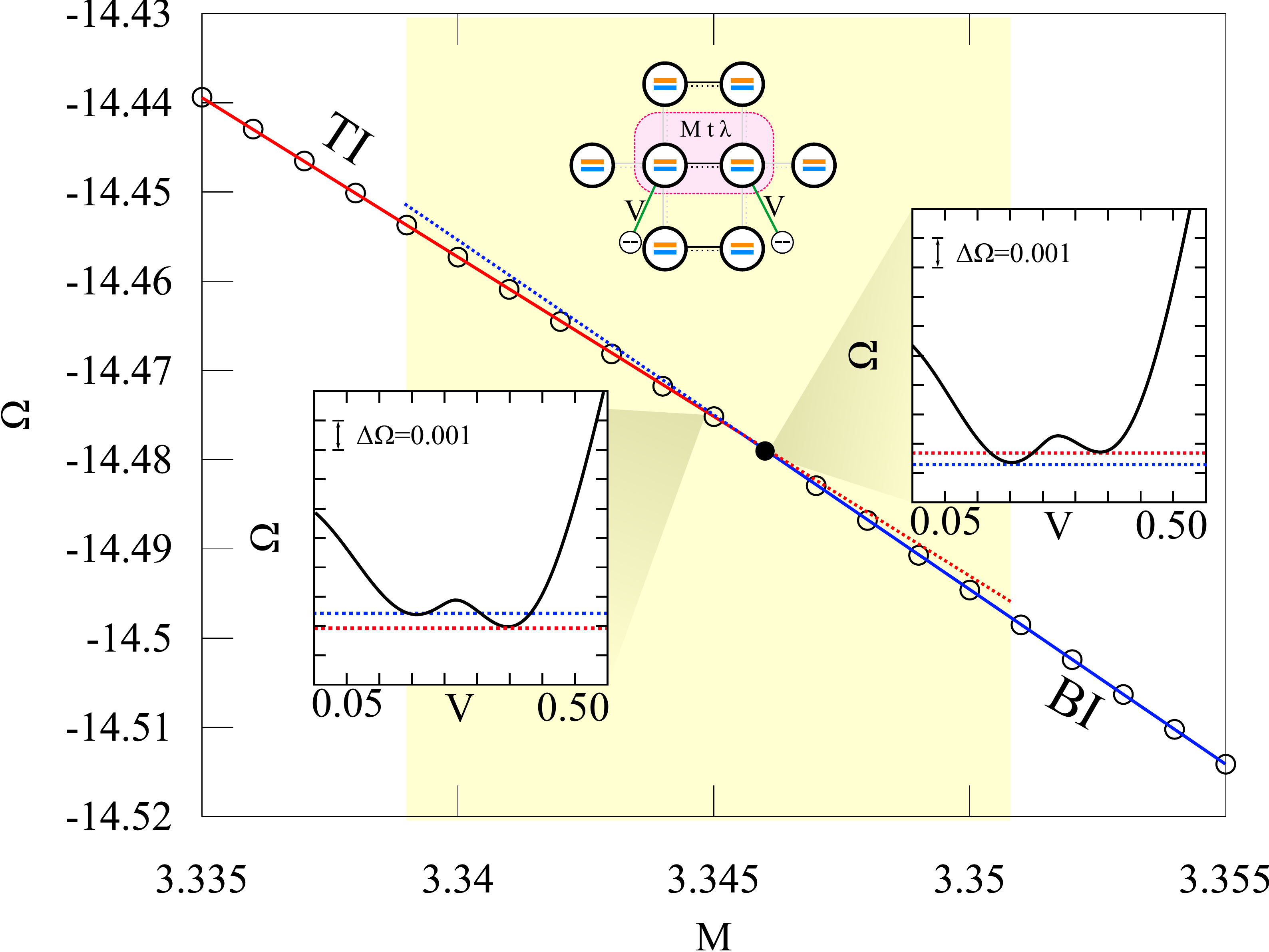}
  \caption{(Color online)
    Internal energy as a function of $M$ for $U=8$ and across the topological transition. The two lines corresponds to solutions obtained started from the QSHI (red) at small $M$ or from the BI (blue) at large $M$. The filled symbol at the crossing indicates the transition point. Data from VCA calculations. The internal energy is the value of the $\Omega$ at the stationary point. Insets: The self-energy functional $\Omega$ as a function of $V$ for $M=3.345$ and $M=3.347$, respectively, before and after the topological transition. The relative position of the minima flips at the transition. Figure adapted from~\cite{CrippaThesis}.
   }
  \label{fig6}
\end{figure}

To futher corroborate the evidence of a first-order transition with a non-local approach, we can study the energetic balance of the different solutions. For this, a variational method like variational cluster approximation (VCA) is particularly suited, and in the remaining of this section we present VCA results.
% for a very similar reference cluster.
%A clear insight on the nature and stability of the discontinuous topological transition occurring at large interaction is given by the behavior of the internal energy presented by the VCA framework. 
%The VCA method offers an alternative route to get the solution of the interacting model. Just like CDMFT, the study of the full problem is reduced to that of a single ``reference" cluster. Yet, unlike in CDMFT,
Here the solutions are obtained by finding the stationary points of a functional of the cluster self-energy $\Omega(\Sigma)$~\cite{Potthoff2003,Potthoff2003PRL,Senechal2010}. The value of the functional in the stationary points corresponds to the grand potential of a physical solution with the corresponding self-energy.
%Different reference clusters can be envisaged for a given problem, and indeed we are free to introduce additional variational parameters in the form, for example, of fictitious external fields or non-interacting bath levels coupled to the interacting cluster.  
VCA requires to use a reference system which we choose in analogy with CDMFT as a two site cluster and four additional bath levels (one per orbital and site) . The local energy of such effective bath levels is fixed at the chemical potential to enforce the half-filling condition. The hybridization $V$ among cluster and bath levels is our variational parameter (see the sketch in  \figu{fig6}).

We can therefore follow the evolution of the internal energy of the two solutions (QSHI and BI) as a function of the mass parameter $M$ ( \figu{fig6}) which clearly shows a wide region of coexistence and a crossing between the two solutions marking the first-order transition.
 The insets show the behavior of $\Omega$ as function of the variational parameter $V$  for two points immediately before and after the TQPT.  In both cases the functional shows two minima, separated by a local maximum.  In the QSHI phase the minimum is the more strongly hybridized one, whereas the BI solution features a weakly coupled state. As the transition is crossed the global maximum jumps from one of the minima to the other in the same points where the energies cross as a function of $M$.
 % The tThe relative energy difference among the two minima changes sign across the transition point, signaled by the crossing of the energy curves.\\
% The results, then, show that the stable solution has a discontinuity in its first derivative at the transition point. 

This scenario is completely different from the non-interacting description of the topological transition, where a single minimum is present for every value of $M$ and it evolves continuously across the QPT, according to the conventional portrait of topological band theory.

%The ``double-well" shape of the grand potential supports the results obtained by the single-site DMFT analysis, suggesting the presence of two competing, topologically different solutions between which the system, driven by the M parameter, discontinuously jumps.

\section{Local and non-local correlations through the phase diagram} \label{sec:phasediag}

\begin{figure}[ht]
  \includegraphics[width=\linewidth]{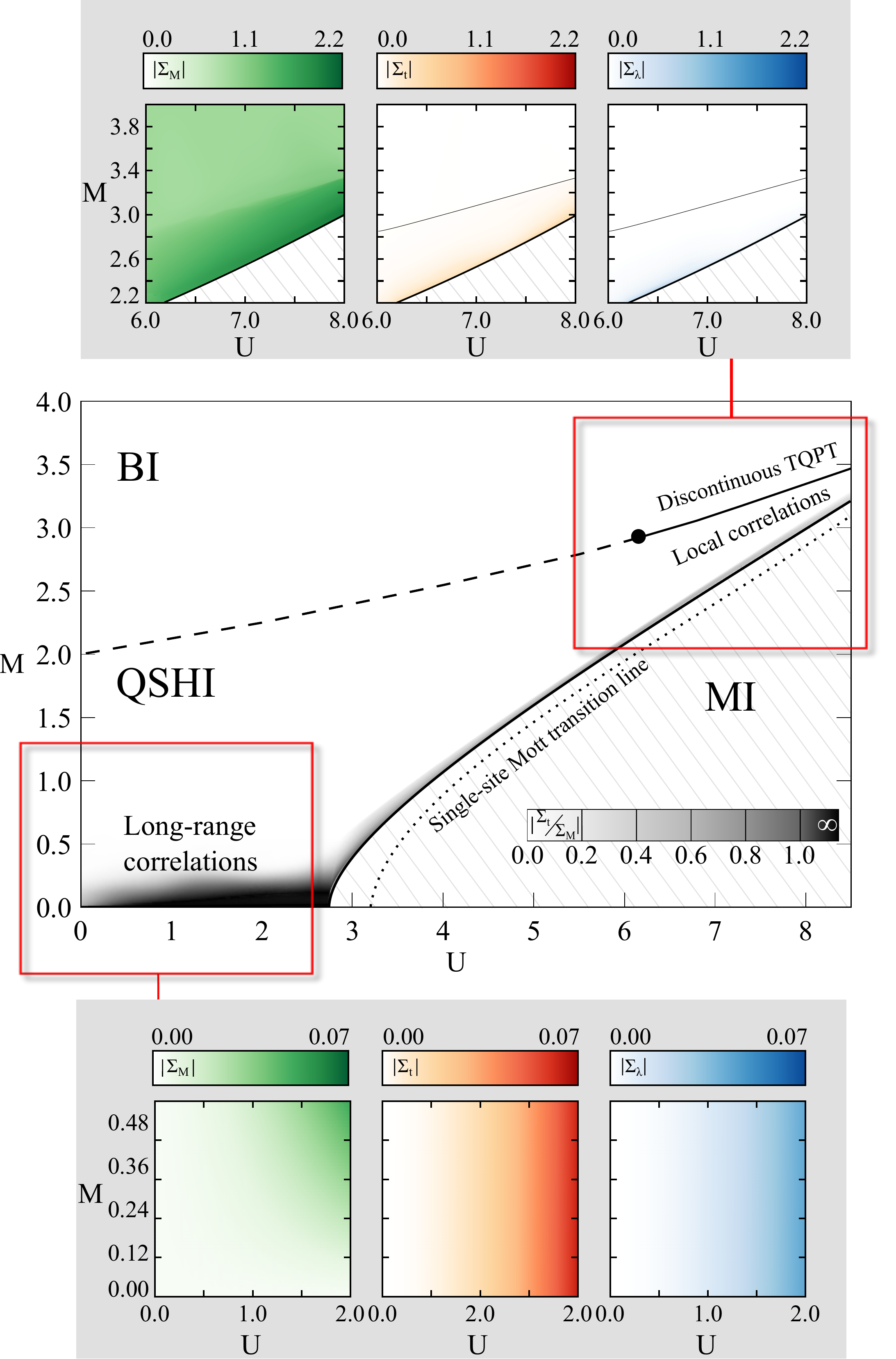}
  \caption{ (Color online) CDMFT phase diagram of the half-filled BHZH model in the $U$-$M$ plane. The three areas refers to band-, quantum spin Hall- and Mott-insulator (BI, QSHI and MI, respectively). In the upper and bottom gradient insets we show the absolute value of the relative $\Sigma$ component at zero frequency. The black gradient in the main panel encodes the relative strength of the momentum-dependent component $\Sigma_{t}$ with respect to $\Sigma_{M}$. The latter represents the dominant type of correlation in the BI and QSHI regions, except close to the Mott line, in particular in the bottom-left corner at small values of $M$. There, even well before the MI phase sets in, the system develops non-local correlations, suggesting the presence of long-range self-energy components.}
  \label{fig7}
\end{figure}

We are now in the position to present a global view of the phase diagram in the $M$-$U$ plane (\figu{fig7}) gathering the information encoded in the cuts that we have been discussing so far.
% summarizes the features of the BHZ-Hubbard system under a cluster-DMFT analysis (VCA results are analogous). The overall shape of the phases is unchanged with respect to the single-particle data 
The transition line between the QSH and BI phases has a positive slope in the $U-M$ parameter plane, accounting for the negative sign of the $\Sigma_{M}$ component, which opposes the level splitting proportional to \textit{M}~\cite{Budich2013PRB,Amaricci2015PRL}. A Mott insulating region with uniform orbital occupancy  opens for large $U$; as a result, the topologically nontrivial phase intrudes between the trivial Mott ($U \gg M$) and band ($U \ll M$) insulator. We mention that the existence of the high-spin Mott insulator follows from the finite Hund's coupling $J$~\cite{Amaricci2016PRB}.
%, favoured at large $U$  and the Band Insulator, favoured at large $M$.\\
%As previously stated, the many-body effects are completely encoded in
%the complex self-energy term. In particular, the positions of the
%poles of the Green's function, related to the single-particle
%excitation spectrum, are controlled by the Hermitian part of the
%self-energy, which couples to the Hamiltonian and shifts its
%eigenvalues.

In the spirit of the present analysis of local and non-local correlations, in the main panel of \figu{fig7} we show the ratio between the non-local and local self-energies $\Sigma_t /\Sigma_M$ in a greyscale. The plot clearly highlights one of the main results emerging from our analysis, namely that the non-local component is comparable with the local one only in a window of small $M$ and in any case before the Mott transition is reached. 
In this region we can expect that longer-range correlations become more and more important. Therefore, in contrast with the large-$M$ scenario, where the self-energy is essentially local, larger clusters would be necessary to accurately describe the system.

The rest of the phase diagram is clearly white -- signaling that local correlations overwhelm non-local ones -- in particular close to the transition line between QSHI and BI. This analysis can not be applied in the Mott insulating phase, where all the self-energy components tend to diverge~\cite{Capone2004}. This is the origin of the very thin grey border at the boundary of the Mott insulating which is in turn
identified with a dashed motif.

The phase diagram shows again that the two regions we have highlighted in the previous sections present a very different competition between local and non-local correlations. In order to highlight this point we show the three components of the self-energy in color scales. The three panels at the bottom of \figu{fig7} illustrate the picture at small $M$. Here we see that the three components evolve differently. While the local $\Sigma_M$ increases when both $U$ and $M$ are simultaneously increased (it is larger in the top-right corner), the non-local components, and in particular $\Sigma_t$, increase with $U$ and they have a weaker dependence on $M$ (they are large moving towards the right of the panels). The balance and the competition between the different components is hence rich and subtle. This can be another indication that this region can be strongly affected by including further non-local effects. 

On the other hand, the large $M$ and large $U$ case (panels on
top of \figu{fig7}) presents a much simpler picture. The local
self-energy is large in the whole QSHI phase, while the two local
components are basically negligible except for a thin parameter range close
to the Mott transition line. In particular,  the
behavior around the topological transition is dominated by $\Sigma_M$
and it can be understood within single-site DMFT.

%\subsection{Comparison with the Kane-Mele-Hubbard model}
%Hence we confirm that, for large values of M and concomitantly large
%U, all the nontrivial physics inherent to the topological phase
%transition is contained in the local component of the
%self-energy. This means that the electronic interaction effects that
%contribute to the emergence of the discontinuous behavior are well
%described by single-site DMFT.\\

The characterization of the relative importance of local and non-local correlation effects on the topological
quantum phase transition in the BHZH model naturally calls for a comparison with 
the KMH model. In the latter, the relevant degrees of freedom for the QSHI state 
are represented by the non-local sub-lattice flavor, rather than by the local orbitals.
The KMH model has been extensively investigated using different
methods, e.g.  VCA, Hartree-Fock, single-site as well as cluster DMFT
~\cite{Rachel2018ROPIP,Rachel2010,Ruegg2012,Zheng2020,Mertz2019,Pizarro2020}. 
% In this section we briefly discuss the relation between the CDMFT
% results presented here for the BHZH and the Kane-Mele-Hubbard (KMH)
% model. Previous works have addressed the latter within various 
% methods, among which VCA, Hartree-Fock, single- as well as CDMFT
% ~\cite{Rachel2010,Ruegg2012,Zheng2020,Mertz2019,Pizarro2020}. 
%Most of
%these studies focused on the effect of electronic correlations onto
%next-nearest neighbor spin-orbit processes or on the transition to
%the antiferromagnetically long-range ordered trivial phase.
%Here, we are interested in comparing the role of the
%intra- and inter-orbital Slater-Kanamori interaction of the BHZH
%model to those of the intra-sublattice electron-electron interaction
%of the KMH.
%The correspondence between the orbital subspace of the BHZH model and the sublattice one of the KMH indicates that 
The local splitting $M$ between the two orbitals in the BHZH model is
replaced, in the KMH, by the staggered potential between two
sublattices of the honeycomb, i.e. the Semenoff mass~\cite{Rachel2018ROPIP} (see \figu{fig8}a). Similarly to the BHZH model, this term
drives the KMH model towards a trivial phase, reached when $M$
becomes larger than a critical value $M_c$~\cite{KaneMele2005}.
To the best of our knowledge, there is no evidence of a discontinuous topological transition in this model
in the presence of large interactions, in contrast with the results of the BHZH model, raising a question
about the role of locality of the interactions as well as of the approximations.

Interestingly, in the  closely related (spinful) Haldane model  it has been shown
by means of a two-particle self-consistent approximation,  
that close to the topological transition
driven by the Semenoff mass at fixed $U$, the dominant renormalization
is represented by the on-site components of the self-energy\cite{Mertz2019}, similarly to what we
found here for the BHZH model.
The combination of these results and those here reported, showing that the discontinuous transition of the BHZH model 
survives the inclusion of non-local effects, suggests that the absence of first-order topological transition in the KMH model 
is not a consequence of the approximation, but it results from the difference between the two many-body Hamiltonians. 

A closer inspection identifies a natural candidate in the interaction terms, which in the BHZH model involves the orbital subspace and in the KMH, the sub-lattice one. 
Indeed, the BHZH features both a (local) intra-orbital repulsion and a (local) inter-orbital one 
whereas the KMH -- at least in its standard form -- has no inter-site term. This breaks the correspondence at the level of the two-body interaction.
%whereas it does not include any inter-site interaction. %it only contains a local $U$.
%The presence of this single interaction scale results in a ``simpler''
%monotonic behavior of $\Sigma_M$ and the multiplet structure is only
%overall affected by $U$.
Indeed, single-site DMFT calculations for the KMH model reported in Appendix {\bf A} confirm the absence 
of a discontinuous transition and, most importantly, they share the
same qualitative behavior of calculations for the BHZH model in which
we artificially switched off the inter-orbital repulsion
~\cite{Budich2013PRB,Amaricci2015PRL,Amaricci2016PRB}. 

Following this analogy we can surmise that the inclusion of non-local
components of the electron-electron interaction would provide a route
to observe first-order effects in the KMH.
Treating the KMH within CDMFT and adding an inter-sublattice interaction mimicking the effect of the Hund's coupling could potentially give information on the small-$M$ non-local physics as well as on the first-order topological phase transition, which instead in the BHZH model is present already at the single-site DMFT level.

%This
%result, generic with respect to $U$, has been tested beyond the DMFT
%description using a two-particle self-consistent
%approximation~\cite{Mertz2019}.
%
%The mechanism is the same discussed here for the BHZH within
%CDMFT: the spatially local self-energy embodies the tendency of the
%Hubbard repulsion to counteract the sublattice/orbital
%polarization. As shown in Fig.~\ref{fig8}(b), $\Sigma_M$
%vanishes for $M\!=\!0$ and, upon increasing $M$, the system is pushed
%from the QSHI phase where the sublattice occupations are more equal,
%towards the trivial insulator with a large staggered potential.

\section{Conclusion}\label{Sec:Conclusions}

We have investigated the electronic correlations throughout the phase diagram of the BHZH model which integrates a typical setup for topological phase transitions and a Hubbard-like local interaction term. 
We use CDMFT, a quantum cluster extension of DMFT that accurately treats short-ranged correlations inside the cluster. 
These are interpreted in terms of additive renormalizations of three control model parameters. In particular we have a local term, which renormalizes the mass $M$ ruling the topological phase transition and two non-local terms correcting the nearest-neighbor hopping $t$ and the spin-orbit coupling $\lambda$, respectively. 

This way, we can show how the relative importance of local and non-local correlations changes in a spectacular way according to the bare parameters and, more precisely, that the value of $M$ has a strong impact. 
The non-local terms turn out to be comparable with the non-local ones for small $M$, as long as the Hubbard $U$ is not sufficient to drive a Mott transition. On the other hand, local correlation effects are predominant at intermediate and large $M$ including the whole region around the topological phase transition. This allows us to confirm the robustness of the scenario derived with single-site DMFT, where the topological transition becomes of first-order for intermediate values of $U$ and $M$. 

The present results are for phases with no long-range order. It is well known that both BHZH and KMH models host lower symmetry magnetic or charge ordered phases, in particular in the correlated context.  
Our findings give indications on the nature of the symmetry-broken phases that may develop in the different parameter ranges.
At large $U$ and $M$, where local self-energy corrections are found to dominate, we expect instabilities associated to the ordering of local spin- or orbital-moments. 
The importance of ${\bf k}$-dependent self-energy corrections at small $M$ suggests instead the possibilily of a wider palette of competing phases. A pronounced itinerant character or the ordering at small momenta may indeed arise close to the orbital-degenerate line $M=0$, where the underlying paramagnetic Mott phase at large $U$ is driven primarily by non-local components of $\Sigma$, rather than by the local ones as in the conventional case. 

%leading to small-$q$ instabilities that are beyond the scope of our cluster methods. 

\section{Acknowledgements}
L.C., A.A. and M.C. acknowledge financial support from  MIUR through PRIN 2015 (Prot. 2015C5SEJJ001) and PRIN2017 project CEnTral (Protocol Number 20172H2SC4). A.A. and M.C. acknowledge support from H2020 Framework Programme, 
under ERC Advanced Grant No. 692670 ``FIRSTORM''.
G.S., S.A. and L.C. acknowledge financial support by the Deutsche Forschungsgemeinschaft (DFG, German Research Foundation) under Germany's Excellence Strategy through W\"urzburg‐Dresden Cluster of Excellence on Complexity and Topology in Quantum Matter ‐ ct.qmat (EXC 2147, project‐id 390858490). The authors gratefully acknowledge the Gauss Centre for Supercomputing e.V. (www.gauss-centre.eu) for funding this project by providing computing time on the GCS Supercomputer SuperMUC at Leibniz Supercomputing Centre (www.lrz.de).

\section*{Appendix A}
%\label{appendix}
The Kane-Mele model is historically the first proposal of a topological insulator presenting a nontrivial quantum spin-Hall  effect~\cite{KaneMele2005}. The setup consists of two copies of the Haldane model on the honeycomb lattice~\cite{Haldane1988}, one for each spin orientation, related by time-reversal symmetry. A Haldane mass term with opposite sign for the two spins is then added: this term, which couples momentum with spin degrees of freedom, arises from spin-orbit coupling. It is also non-local, taking the form of a second-nearest-neighbour hopping amplitude. In contrast with the Haldane model, the Kane-Mele mass term does not break the time-reversal symmetry of the system.\\

\begin{figure}[h]
  \includegraphics[width=\linewidth]{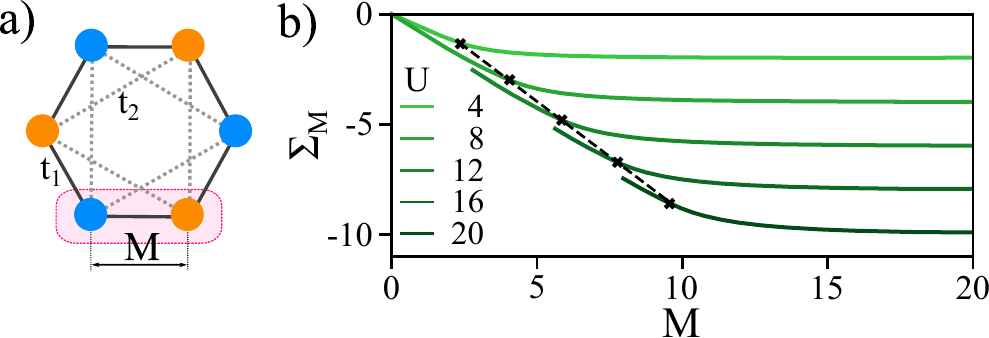}
  \caption{(Color online)
  \textbf{a)} Schematics of the Kane-Mele model on the honeycomb lattice.
  \textbf{b)} Single-site DMFT data for site-diagonal self-energy component $\Sigma_M$ in the KMH as a function of $M$ for different values of $U$. Every line is continuous independently on how large $U$ is chosen, denoting the absence of the first-order phase transition. This is at odds with the BHZ model at large $M$ and large $U$ where both single-site and cluster DMFT show a jump in $\Sigma_M$ (see \figu{fig5}). The black crosses show the location of the topological phase transition in the KMH model.
   }
  \label{fig8}
\end{figure}

In reciprocal space, the Hamiltonian matrix  of the model has the form 
\begin{equation}
H(\textbf{k})=\sum_{a=1}^{5}d_{a}(\mathbf{k})\Gamma^{a}+\sum_{a<b=1}^{5}d_{ab}(\mathbf{k})\Gamma^{ab},
\end{equation}
where the $d$ terms are trigonometric functions of the momentum components, and the $\Gamma$ matrices span the 16 generators of the $SU(4)$ matrix group (excluding the identity)~\cite{KaneMele2005b}: in particular, the five $\Gamma^{a}$ are the so-called Dirac matrices satisfying the Clifford algebra
$$
[\Gamma^{\mu},\Gamma^{\nu}]_{+}=2\delta_{\mu\nu}\mathbb{I}_{4\times4}
$$
while the remaining $\Gamma^{ab}$ are obtained from the commutators of the Dirac matrices,
$$
\Gamma^{\mu\nu}=\dfrac{1}{2i}[\Gamma^{\mu},\Gamma^{\nu}].
$$

The effect of the single-site DMFT self-energy on the topology of the KMH model can be evinced from~\figu{fig8}. As established in the previous sections, the discontinuity in the topological phase transition of the BHZ model is due to the local correlation effects, i.e. to the real part of the local self-energy component $\Sigma_M$. Its behavior in the single-orbital KMH model is shown, for various values of the Hubbard interaction, in panel (b), with the topological phase transition points for the different $M$ values being marked by the black crosses.
%Here, we look at the Kane-Mele model, i.e. we do not break time-reversal invariance, and focus on the BI-QSHI line at large values of $M$. 
%In this region we know, thanks to our analysis of the BHZH model presented in the previous section, that we can safely rely on single-site DMFT.  
%In Fig.~\ref{fig8} we plot $\Sigma_M$ across the topological phase transition for various values of $U$ and observe several similarities and one difference w.r.t. the BHZH:
%As in the BI phase of the BHZH, the local component of the self-energy hardly depends on $M$ at large values of $M$, i.e. in the trivial phase, as the system is a fully polarized (among sublattices in the KM and among orbitals in the BHZ) band insulator. 
%Upon lowering $M$, the system reaches the point at which the sublattice character gets inverted at the valley momenta (in the BHZ model the orbital character at the $\Gamma$ point gets instead inverted). Here, the $\mathbb{Z}_2$-invariant changes from 0 to 1 and, concomitantly, the sublattice/orbital polarization starts to be different from zero.
%At the same time, the local component of $\Sigma_M$ starts to be active in contrasting the sublattice/orbital mixing or fluctuations, which means that $\Sigma_M$ is no longer flat as a function of $M$, see Fig.~\ref{fig8}b. 
\begin{figure}[h]
  \includegraphics[width=\linewidth]{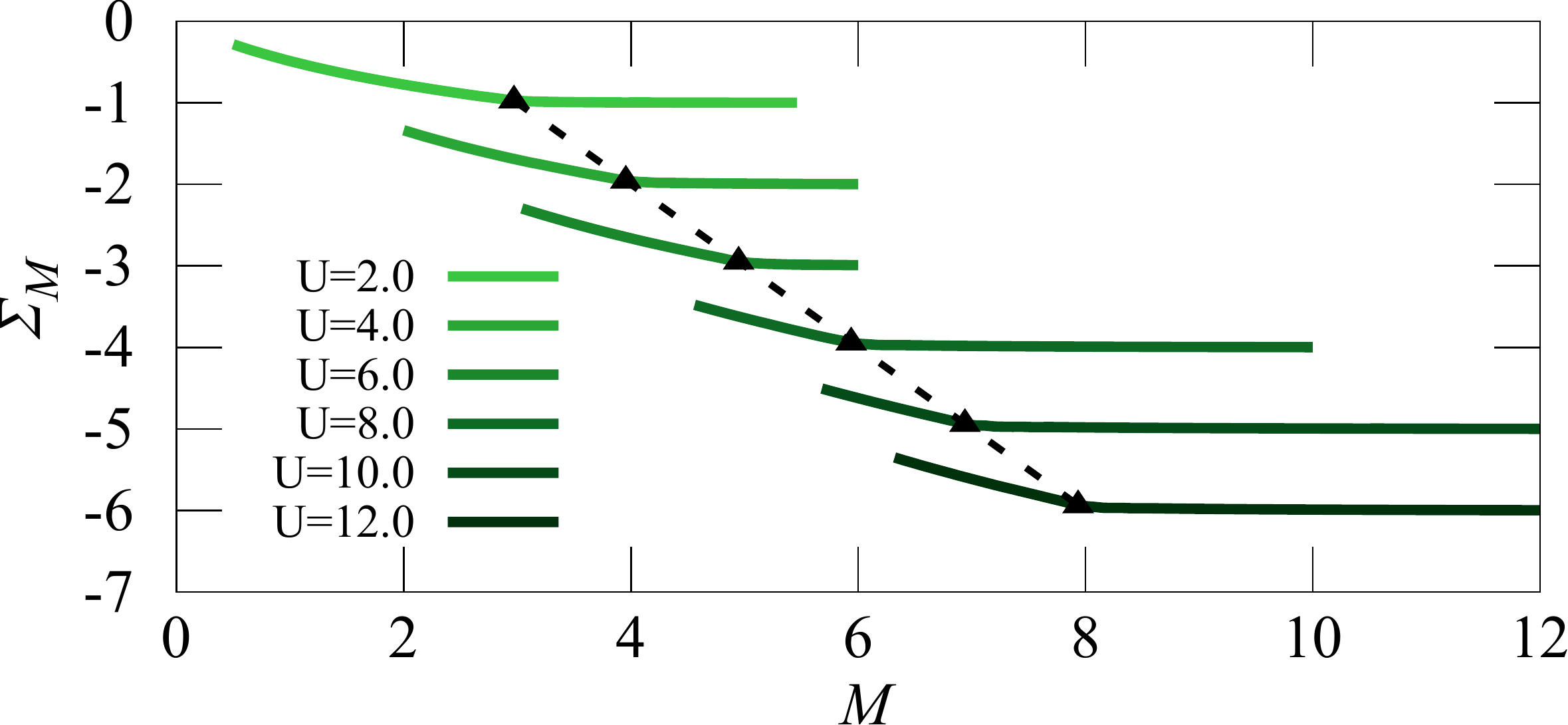} 
  \caption{(Color online) Real part of the self-energy for the BHZ model where the interaction term cointains no inter-orbital component. The data refer to single-site DMFT, where only the diagonal component is present. The black marks and dotted line refer to the values of $\Sigma_{M}$ where the TQPT occurs, the latter being always continuous. As in \figu{fig8}, and in contrast to \figu{fig5}, the topological phase transition is always continuous, since $\Sigma_{M}$ features no discontinuities. }
\label{fig9}
\end{figure}

The comparison of \figu{fig8}(b) on the one side and \figu{fig3} and \figu{fig5} on the other evidences two main differences:
$i$) the minimum of $\Sigma_M$ at intermediate values of $M$ reported
in \figu{fig3}(a) and (b) - which, crucially, characterizes both DMFT and CDMFT results - is absent for the KMH case, and $ii$) the jump at large values of $U$, visible in \figu{fig5}, is not present. 
Both points can be linked to the difference in the internal structure
of the two-body interaction Hamiltonian. This can be shown by comparing these results with a BHZH setup in which $H_{\mathrm{int}}$ consists of the orbital-diagonal Hubbard U term only, disregarding all effects of Hund's coupling. This setup is shown in \figu{fig9}, where the values of $\Sigma_{M}$ are again plotted as a function of the bare mass for various interaction strengths. The absence of discontinuities in the self-energy evolution, which reflects in a continuous topological phase transition in the $U-M$ phase diagram, suggests that the orbital off-diagonal terms of the Slater-Kanamori interaction are the main drivers of the discontinuous TQPT.

\bibliography{references}

\begin{thebibliography}{70}%
\makeatletter
\providecommand \@ifxundefined [1]{%
 \@ifx{#1\undefined}
}%
\providecommand \@ifnum [1]{%
 \ifnum #1\expandafter \@firstoftwo
 \else \expandafter \@secondoftwo
 \fi
}%
\providecommand \@ifx [1]{%
 \ifx #1\expandafter \@firstoftwo
 \else \expandafter \@secondoftwo
 \fi
}%
\providecommand \natexlab [1]{#1}%
\providecommand \enquote  [1]{``#1''}%
\providecommand \bibnamefont  [1]{#1}%
\providecommand \bibfnamefont [1]{#1}%
\providecommand \citenamefont [1]{#1}%
\providecommand \href@noop [0]{\@secondoftwo}%
\providecommand \href [0]{\begingroup \@sanitize@url \@href}%
\providecommand \@href[1]{\@@startlink{#1}\@@href}%
\providecommand \@@href[1]{\endgroup#1\@@endlink}%
\providecommand \@sanitize@url [0]{\catcode `\\12\catcode `\$12\catcode
  `\&12\catcode `\#12\catcode `\^12\catcode `\_12\catcode `\%12\relax}%
\providecommand \@@startlink[1]{}%
\providecommand \@@endlink[0]{}%
\providecommand \url  [0]{\begingroup\@sanitize@url \@url }%
\providecommand \@url [1]{\endgroup\@href {#1}{\urlprefix }}%
\providecommand \urlprefix  [0]{URL }%
\providecommand \Eprint [0]{\href }%
\providecommand \doibase [0]{https://doi.org/}%
\providecommand \selectlanguage [0]{\@gobble}%
\providecommand \bibinfo  [0]{\@secondoftwo}%
\providecommand \bibfield  [0]{\@secondoftwo}%
\providecommand \translation [1]{[#1]}%
\providecommand \BibitemOpen [0]{}%
\providecommand \bibitemStop [0]{}%
\providecommand \bibitemNoStop [0]{.\EOS\space}%
\providecommand \EOS [0]{\spacefactor3000\relax}%
\providecommand \BibitemShut  [1]{\csname bibitem#1\endcsname}%
\let\auto@bib@innerbib\@empty
%</preamble>
\bibitem [{\citenamefont {Bloch}(1929)}]{Bloch1929}%
  \BibitemOpen
  \bibfield  {author} {\bibinfo {author} {\bibfnamefont {F.}~\bibnamefont
  {Bloch}},\ }\bibfield  {title} {\bibinfo {title} {Über die quantenmechanik
  der elektronen in kristallgittern},\ }\href
  {https://doi.org/10.1007/BF01339455} {\bibfield  {journal} {\bibinfo
  {journal} {Zeitschrift für Physik}\ }\textbf {\bibinfo {volume} {52}},\
  \bibinfo {pages} {555} (\bibinfo {year} {1929})}\BibitemShut {NoStop}%
\bibitem [{\citenamefont {Dai}\ \emph {et~al.}(2008)\citenamefont {Dai},
  \citenamefont {Hughes}, \citenamefont {Qi}, \citenamefont {Fang},\ and\
  \citenamefont {Zhang}}]{Dai2008}%
  \BibitemOpen
  \bibfield  {author} {\bibinfo {author} {\bibfnamefont {X.}~\bibnamefont
  {Dai}}, \bibinfo {author} {\bibfnamefont {T.~L.}\ \bibnamefont {Hughes}},
  \bibinfo {author} {\bibfnamefont {X.-L.}\ \bibnamefont {Qi}}, \bibinfo
  {author} {\bibfnamefont {Z.}~\bibnamefont {Fang}},\ and\ \bibinfo {author}
  {\bibfnamefont {S.-C.}\ \bibnamefont {Zhang}},\ }\bibfield  {title} {\bibinfo
  {title} {Helical edge and surface states in hgte quantum wells and bulk
  insulators},\ }\href {https://doi.org/10.1103/PhysRevB.77.125319} {\bibfield
  {journal} {\bibinfo  {journal} {Phys. Rev. B}\ }\textbf {\bibinfo {volume}
  {77}},\ \bibinfo {pages} {125319} (\bibinfo {year} {2008})}\BibitemShut
  {NoStop}%
\bibitem [{\citenamefont {Br\"une}\ \emph {et~al.}(2011)\citenamefont
  {Br\"une}, \citenamefont {Liu}, \citenamefont {Novik}, \citenamefont
  {Hankiewicz}, \citenamefont {Buhmann}, \citenamefont {Chen}, \citenamefont
  {Qi}, \citenamefont {Shen}, \citenamefont {Zhang},\ and\ \citenamefont
  {Molenkamp}}]{Brune2011}%
  \BibitemOpen
  \bibfield  {author} {\bibinfo {author} {\bibfnamefont {C.}~\bibnamefont
  {Br\"une}}, \bibinfo {author} {\bibfnamefont {C.~X.}\ \bibnamefont {Liu}},
  \bibinfo {author} {\bibfnamefont {E.~G.}\ \bibnamefont {Novik}}, \bibinfo
  {author} {\bibfnamefont {E.~M.}\ \bibnamefont {Hankiewicz}}, \bibinfo
  {author} {\bibfnamefont {H.}~\bibnamefont {Buhmann}}, \bibinfo {author}
  {\bibfnamefont {Y.~L.}\ \bibnamefont {Chen}}, \bibinfo {author}
  {\bibfnamefont {X.~L.}\ \bibnamefont {Qi}}, \bibinfo {author} {\bibfnamefont
  {Z.~X.}\ \bibnamefont {Shen}}, \bibinfo {author} {\bibfnamefont {S.~C.}\
  \bibnamefont {Zhang}},\ and\ \bibinfo {author} {\bibfnamefont {L.~W.}\
  \bibnamefont {Molenkamp}},\ }\bibfield  {title} {\bibinfo {title} {Quantum
  hall effect from the topological surface states of strained bulk hgte},\
  }\href {https://doi.org/10.1103/PhysRevLett.106.126803} {\bibfield  {journal}
  {\bibinfo  {journal} {Phys. Rev. Lett.}\ }\textbf {\bibinfo {volume} {106}},\
  \bibinfo {pages} {126803} (\bibinfo {year} {2011})}\BibitemShut {NoStop}%
\bibitem [{\citenamefont {Bernevig}\ \emph {et~al.}(2006)\citenamefont
  {Bernevig}, \citenamefont {Hughes},\ and\ \citenamefont {Zhang}}]{BHZ2006}%
  \BibitemOpen
  \bibfield  {author} {\bibinfo {author} {\bibfnamefont {B.~A.}\ \bibnamefont
  {Bernevig}}, \bibinfo {author} {\bibfnamefont {T.~L.}\ \bibnamefont
  {Hughes}},\ and\ \bibinfo {author} {\bibfnamefont {S.-C.}\ \bibnamefont
  {Zhang}},\ }\bibfield  {title} {\bibinfo {title} {Quantum spin hall effect
  and topological phase transition in hgte quantum wells},\ }\href
  {https://doi.org/10.1126/science.1133734} {\bibfield  {journal} {\bibinfo
  {journal} {Science}\ }\textbf {\bibinfo {volume} {314}},\ \bibinfo {pages}
  {1757} (\bibinfo {year} {2006})}\BibitemShut {NoStop}%
\bibitem [{\citenamefont {Rothe}\ \emph {et~al.}(2010)\citenamefont {Rothe},
  \citenamefont {Reinthaler}, \citenamefont {Liu}, \citenamefont {Molenkamp},
  \citenamefont {Zhang},\ and\ \citenamefont {Hankiewicz}}]{Rothe2010}%
  \BibitemOpen
  \bibfield  {author} {\bibinfo {author} {\bibfnamefont {D.~G.}\ \bibnamefont
  {Rothe}}, \bibinfo {author} {\bibfnamefont {R.~W.}\ \bibnamefont
  {Reinthaler}}, \bibinfo {author} {\bibfnamefont {C.-X.}\ \bibnamefont {Liu}},
  \bibinfo {author} {\bibfnamefont {L.~W.}\ \bibnamefont {Molenkamp}}, \bibinfo
  {author} {\bibfnamefont {S.-C.}\ \bibnamefont {Zhang}},\ and\ \bibinfo
  {author} {\bibfnamefont {E.~M.}\ \bibnamefont {Hankiewicz}},\ }\bibfield
  {title} {\bibinfo {title} {Fingerprint of different spin{\textendash}orbit
  terms for spin transport in {HgTe} quantum wells},\ }\href
  {https://doi.org/10.1088/1367-2630/12/6/065012} {\bibfield  {journal}
  {\bibinfo  {journal} {New Journal of Physics}\ }\textbf {\bibinfo {volume}
  {12}},\ \bibinfo {pages} {065012} (\bibinfo {year} {2010})}\BibitemShut
  {NoStop}%
\bibitem [{\citenamefont {Kane}\ and\ \citenamefont
  {Mele}(2005{\natexlab{a}})}]{Kane2005PRLa}%
  \BibitemOpen
  \bibfield  {author} {\bibinfo {author} {\bibfnamefont {C.~L.}\ \bibnamefont
  {Kane}}\ and\ \bibinfo {author} {\bibfnamefont {E.~J.}\ \bibnamefont
  {Mele}},\ }\bibfield  {title} {\bibinfo {title} {{Quantum Spin Hall Effect in
  Graphene}},\ }\href {https://doi.org/10.1103/PhysRevLett.95.226801}
  {\bibfield  {journal} {\bibinfo  {journal} {Phys. Rev. Lett.}\ }\textbf
  {\bibinfo {volume} {95}},\ \bibinfo {pages} {226801} (\bibinfo {year}
  {2005}{\natexlab{a}})}\BibitemShut {NoStop}%
\bibitem [{\citenamefont {Kane}\ and\ \citenamefont
  {Mele}(2005{\natexlab{b}})}]{Kane2005PRL}%
  \BibitemOpen
  \bibfield  {author} {\bibinfo {author} {\bibfnamefont {C.~L.}\ \bibnamefont
  {Kane}}\ and\ \bibinfo {author} {\bibfnamefont {E.~J.}\ \bibnamefont
  {Mele}},\ }\bibfield  {title} {\bibinfo {title} {{${Z}_{2}$ Topological Order
  and the Quantum Spin Hall Effect}},\ }\href
  {https://doi.org/10.1103/PhysRevLett.95.146802} {\bibfield  {journal}
  {\bibinfo  {journal} {Phys. Rev. Lett.}\ }\textbf {\bibinfo {volume} {95}},\
  \bibinfo {pages} {146802} (\bibinfo {year} {2005}{\natexlab{b}})}\BibitemShut
  {NoStop}%
\bibitem [{\citenamefont {Fu}\ and\ \citenamefont {Kane}(2007)}]{FuKane2007}%
  \BibitemOpen
  \bibfield  {author} {\bibinfo {author} {\bibfnamefont {L.}~\bibnamefont
  {Fu}}\ and\ \bibinfo {author} {\bibfnamefont {C.~L.}\ \bibnamefont {Kane}},\
  }\bibfield  {title} {\bibinfo {title} {Topological insulators with inversion
  symmetry},\ }\href {https://doi.org/10.1103/PhysRevB.76.045302} {\bibfield
  {journal} {\bibinfo  {journal} {Phys. Rev. B}\ }\textbf {\bibinfo {volume}
  {76}},\ \bibinfo {pages} {045302} (\bibinfo {year} {2007})}\BibitemShut
  {NoStop}%
\bibitem [{\citenamefont {Kane}(2008)}]{Kane2008NP}%
  \BibitemOpen
  \bibfield  {author} {\bibinfo {author} {\bibfnamefont {C.~L.}\ \bibnamefont
  {Kane}},\ }\bibfield  {title} {\bibinfo {title} {{Condensed matter: An
  insulator with a twist}},\ }\href {http://dx.doi.org/10.1038/nphys955}
  {\bibfield  {journal} {\bibinfo  {journal} {Nat Phys}\ }\textbf {\bibinfo
  {volume} {4}},\ \bibinfo {pages} {348} (\bibinfo {year} {2008})}\BibitemShut
  {NoStop}%
\bibitem [{\citenamefont {Hasan}\ and\ \citenamefont
  {Kane}(2010)}]{Hasan2010RMP}%
  \BibitemOpen
  \bibfield  {author} {\bibinfo {author} {\bibfnamefont {M.~Z.}\ \bibnamefont
  {Hasan}}\ and\ \bibinfo {author} {\bibfnamefont {C.~L.}\ \bibnamefont
  {Kane}},\ }\bibfield  {title} {\bibinfo {title} {Colloquium: Topological
  insulators},\ }\href {https://doi.org/10.1103/RevModPhys.82.3045} {\bibfield
  {journal} {\bibinfo  {journal} {Rev. Mod. Phys.}\ }\textbf {\bibinfo {volume}
  {82}},\ \bibinfo {pages} {3045} (\bibinfo {year} {2010})}\BibitemShut
  {NoStop}%
\bibitem [{\citenamefont {Abrikosov}\ and\ \citenamefont
  {Khalatnikov}(1959)}]{Abrikosov1959}%
  \BibitemOpen
  \bibfield  {author} {\bibinfo {author} {\bibfnamefont {A.~A.}\ \bibnamefont
  {Abrikosov}}\ and\ \bibinfo {author} {\bibfnamefont {I.~M.}\ \bibnamefont
  {Khalatnikov}},\ }\bibfield  {title} {\bibinfo {title} {The theory of a fermi
  liquid (the properties of liquid 3he at low temperatures)},\ }\href
  {https://doi.org/10.1088/0034-4885/22/1/310} {\bibfield  {journal} {\bibinfo
  {journal} {Reports on Progress in Physics}\ }\textbf {\bibinfo {volume}
  {22}},\ \bibinfo {pages} {329} (\bibinfo {year} {1959})}\BibitemShut
  {NoStop}%
\bibitem [{\citenamefont {Hubbard}\ and\ \citenamefont
  {Flowers}(1963)}]{Hubbard1}%
  \BibitemOpen
  \bibfield  {author} {\bibinfo {author} {\bibfnamefont {J.}~\bibnamefont
  {Hubbard}}\ and\ \bibinfo {author} {\bibfnamefont {B.~H.}\ \bibnamefont
  {Flowers}},\ }\bibfield  {title} {\bibinfo {title} {Electron correlations in
  narrow energy bands},\ }\href {https://doi.org/10.1098/rspa.1963.0204}
  {\bibfield  {journal} {\bibinfo  {journal} {Proceedings of the Royal Society
  of London. Series A. Mathematical and Physical Sciences}\ }\textbf {\bibinfo
  {volume} {276}},\ \bibinfo {pages} {238} (\bibinfo {year}
  {1963})}\BibitemShut {NoStop}%
\bibitem [{\citenamefont {Hubbard}\ and\ \citenamefont
  {Flowers}(1964{\natexlab{a}})}]{Hubbard2}%
  \BibitemOpen
  \bibfield  {author} {\bibinfo {author} {\bibfnamefont {J.}~\bibnamefont
  {Hubbard}}\ and\ \bibinfo {author} {\bibfnamefont {B.~H.}\ \bibnamefont
  {Flowers}},\ }\bibfield  {title} {\bibinfo {title} {Electron correlations in
  narrow energy bands. ii. the degenerate band case},\ }\href
  {https://doi.org/10.1098/rspa.1964.0019} {\bibfield  {journal} {\bibinfo
  {journal} {Proceedings of the Royal Society of London. Series A. Mathematical
  and Physical Sciences}\ }\textbf {\bibinfo {volume} {277}},\ \bibinfo {pages}
  {237} (\bibinfo {year} {1964}{\natexlab{a}})}\BibitemShut {NoStop}%
\bibitem [{\citenamefont {Hubbard}\ and\ \citenamefont
  {Flowers}(1964{\natexlab{b}})}]{Hubbard3}%
  \BibitemOpen
  \bibfield  {author} {\bibinfo {author} {\bibfnamefont {J.}~\bibnamefont
  {Hubbard}}\ and\ \bibinfo {author} {\bibfnamefont {B.~H.}\ \bibnamefont
  {Flowers}},\ }\bibfield  {title} {\bibinfo {title} {Electron correlations in
  narrow energy bands iii. an improved solution},\ }\href
  {https://doi.org/10.1098/rspa.1964.0190} {\bibfield  {journal} {\bibinfo
  {journal} {Proceedings of the Royal Society of London. Series A. Mathematical
  and Physical Sciences}\ }\textbf {\bibinfo {volume} {281}},\ \bibinfo {pages}
  {401} (\bibinfo {year} {1964}{\natexlab{b}})}\BibitemShut {NoStop}%
\bibitem [{\citenamefont {Mott}(1949)}]{Mott1949}%
  \BibitemOpen
  \bibfield  {author} {\bibinfo {author} {\bibfnamefont {N.~F.}\ \bibnamefont
  {Mott}},\ }\bibfield  {title} {\bibinfo {title} {The basis of the electron
  theory of metals, with special reference to the transition metals},\ }\href
  {https://doi.org/10.1088/0370-1298/62/7/303} {\bibfield  {journal} {\bibinfo
  {journal} {Proceedings of the Physical Society. Section A}\ }\textbf
  {\bibinfo {volume} {62}},\ \bibinfo {pages} {416} (\bibinfo {year}
  {1949})}\BibitemShut {NoStop}%
\bibitem [{\citenamefont {Kotliar}\ and\ \citenamefont
  {Vollhardt}(2004)}]{Kotliar2004PT}%
  \BibitemOpen
  \bibfield  {author} {\bibinfo {author} {\bibfnamefont {G.}~\bibnamefont
  {Kotliar}}\ and\ \bibinfo {author} {\bibfnamefont {D.}~\bibnamefont
  {Vollhardt}},\ }\bibfield  {title} {\bibinfo {title} {Strongly correlated
  materials: Insights from dynamical mean-field theory},\ }\href
  {https://doi.org/10.1063/1.1712502} {\bibfield  {journal} {\bibinfo
  {journal} {Physics Today}\ }\textbf {\bibinfo {volume} {57}},\ \bibinfo
  {pages} {53} (\bibinfo {year} {2004})}\BibitemShut {NoStop}%
\bibitem [{\citenamefont {Georges}\ \emph {et~al.}(1996)\citenamefont
  {Georges}, \citenamefont {Kotliar}, \citenamefont {Krauth},\ and\
  \citenamefont {Rozenberg}}]{GeorgesKotliar1996}%
  \BibitemOpen
  \bibfield  {author} {\bibinfo {author} {\bibfnamefont {A.}~\bibnamefont
  {Georges}}, \bibinfo {author} {\bibfnamefont {G.}~\bibnamefont {Kotliar}},
  \bibinfo {author} {\bibfnamefont {W.}~\bibnamefont {Krauth}},\ and\ \bibinfo
  {author} {\bibfnamefont {M.~J.}\ \bibnamefont {Rozenberg}},\ }\bibfield
  {title} {\bibinfo {title} {Dynamical mean-field theory of strongly correlated
  fermion systems and the limit of infinite dimensions},\ }\href
  {https://doi.org/10.1103/RevModPhys.68.13} {\bibfield  {journal} {\bibinfo
  {journal} {Rev. Mod. Phys.}\ }\textbf {\bibinfo {volume} {68}},\ \bibinfo
  {pages} {13} (\bibinfo {year} {1996})}\BibitemShut {NoStop}%
\bibitem [{\citenamefont {Capone}\ \emph {et~al.}(2004)\citenamefont {Capone},
  \citenamefont {Civelli}, \citenamefont {Kancharla}, \citenamefont
  {Castellani},\ and\ \citenamefont {Kotliar}}]{Capone2004}%
  \BibitemOpen
  \bibfield  {author} {\bibinfo {author} {\bibfnamefont {M.}~\bibnamefont
  {Capone}}, \bibinfo {author} {\bibfnamefont {M.}~\bibnamefont {Civelli}},
  \bibinfo {author} {\bibfnamefont {S.~S.}\ \bibnamefont {Kancharla}}, \bibinfo
  {author} {\bibfnamefont {C.}~\bibnamefont {Castellani}},\ and\ \bibinfo
  {author} {\bibfnamefont {G.}~\bibnamefont {Kotliar}},\ }\bibfield  {title}
  {\bibinfo {title} {Cluster-dynamical mean-field theory of the density-driven
  mott transition in the one-dimensional hubbard model},\ }\href
  {https://doi.org/10.1103/PhysRevB.69.195105} {\bibfield  {journal} {\bibinfo
  {journal} {Phys. Rev. B}\ }\textbf {\bibinfo {volume} {69}},\ \bibinfo
  {pages} {195105} (\bibinfo {year} {2004})}\BibitemShut {NoStop}%
\bibitem [{\citenamefont {Park}\ \emph {et~al.}(2008)\citenamefont {Park},
  \citenamefont {Haule},\ and\ \citenamefont {Kotliar}}]{Park2008}%
  \BibitemOpen
  \bibfield  {author} {\bibinfo {author} {\bibfnamefont {H.}~\bibnamefont
  {Park}}, \bibinfo {author} {\bibfnamefont {K.}~\bibnamefont {Haule}},\ and\
  \bibinfo {author} {\bibfnamefont {G.}~\bibnamefont {Kotliar}},\ }\bibfield
  {title} {\bibinfo {title} {Cluster dynamical mean field theory of the mott
  transition},\ }\href {https://doi.org/10.1103/PhysRevLett.101.186403}
  {\bibfield  {journal} {\bibinfo  {journal} {Phys. Rev. Lett.}\ }\textbf
  {\bibinfo {volume} {101}},\ \bibinfo {pages} {186403} (\bibinfo {year}
  {2008})}\BibitemShut {NoStop}%
\bibitem [{\citenamefont {Sakai}\ \emph {et~al.}(2012)\citenamefont {Sakai},
  \citenamefont {Sangiovanni}, \citenamefont {Civelli}, \citenamefont {Motome},
  \citenamefont {Held},\ and\ \citenamefont {Imada}}]{Sakai2012}%
  \BibitemOpen
  \bibfield  {author} {\bibinfo {author} {\bibfnamefont {S.}~\bibnamefont
  {Sakai}}, \bibinfo {author} {\bibfnamefont {G.}~\bibnamefont {Sangiovanni}},
  \bibinfo {author} {\bibfnamefont {M.}~\bibnamefont {Civelli}}, \bibinfo
  {author} {\bibfnamefont {Y.}~\bibnamefont {Motome}}, \bibinfo {author}
  {\bibfnamefont {K.}~\bibnamefont {Held}},\ and\ \bibinfo {author}
  {\bibfnamefont {M.}~\bibnamefont {Imada}},\ }\bibfield  {title} {\bibinfo
  {title} {Cluster-size dependence in cellular dynamical mean-field theory},\
  }\href {https://doi.org/10.1103/PhysRevB.85.035102} {\bibfield  {journal}
  {\bibinfo  {journal} {Phys. Rev. B}\ }\textbf {\bibinfo {volume} {85}},\
  \bibinfo {pages} {035102} (\bibinfo {year} {2012})}\BibitemShut {NoStop}%
\bibitem [{\citenamefont {Potthoff}\ \emph {et~al.}(2003)\citenamefont
  {Potthoff}, \citenamefont {Aichhorn},\ and\ \citenamefont
  {Dahnken}}]{Potthoff2003PRL}%
  \BibitemOpen
  \bibfield  {author} {\bibinfo {author} {\bibfnamefont {M.}~\bibnamefont
  {Potthoff}}, \bibinfo {author} {\bibfnamefont {M.}~\bibnamefont {Aichhorn}},\
  and\ \bibinfo {author} {\bibfnamefont {C.}~\bibnamefont {Dahnken}},\
  }\bibfield  {title} {\bibinfo {title} {Variational cluster approach to
  correlated electron systems in low dimensions},\ }\href
  {https://doi.org/10.1103/PhysRevLett.91.206402} {\bibfield  {journal}
  {\bibinfo  {journal} {Phys. Rev. Lett.}\ }\textbf {\bibinfo {volume} {91}},\
  \bibinfo {pages} {206402} (\bibinfo {year} {2003})}\BibitemShut {NoStop}%
\bibitem [{\citenamefont {Aichhorn}\ \emph {et~al.}(2006)\citenamefont
  {Aichhorn}, \citenamefont {Arrigoni}, \citenamefont {Potthoff},\ and\
  \citenamefont {Hanke}}]{Aichhorn2006}%
  \BibitemOpen
  \bibfield  {author} {\bibinfo {author} {\bibfnamefont {M.}~\bibnamefont
  {Aichhorn}}, \bibinfo {author} {\bibfnamefont {E.}~\bibnamefont {Arrigoni}},
  \bibinfo {author} {\bibfnamefont {M.}~\bibnamefont {Potthoff}},\ and\
  \bibinfo {author} {\bibfnamefont {W.}~\bibnamefont {Hanke}},\ }\bibfield
  {title} {\bibinfo {title} {Variational cluster approach to the hubbard model:
  Phase-separation tendency and finite-size effects},\ }\href
  {https://doi.org/10.1103/PhysRevB.74.235117} {\bibfield  {journal} {\bibinfo
  {journal} {Phys. Rev. B}\ }\textbf {\bibinfo {volume} {74}},\ \bibinfo
  {pages} {235117} (\bibinfo {year} {2006})}\BibitemShut {NoStop}%
\bibitem [{\citenamefont {Sénéchal}(2008)}]{Senechal2008}%
  \BibitemOpen
  \bibfield  {author} {\bibinfo {author} {\bibfnamefont {D.}~\bibnamefont
  {Sénéchal}},\ }\bibfield  {title} {\bibinfo {title} {The variational
  cluster approximation for hubbard models: Practical implementation},\ }in\
  \href {https://doi.org/10.1109/HPCS.2008.18} {\emph {\bibinfo {booktitle}
  {2008 22nd International Symposium on High Performance Computing Systems and
  Applications}}}\ (\bibinfo {year} {2008})\ pp.\ \bibinfo {pages}
  {9--15}\BibitemShut {NoStop}%
\bibitem [{\citenamefont {Balzer}\ \emph {et~al.}(2009)\citenamefont {Balzer},
  \citenamefont {Kyung}, \citenamefont {S{\'{e}}n{\'{e}}chal}, \citenamefont
  {Tremblay},\ and\ \citenamefont {Potthoff}}]{Balzer2009}%
  \BibitemOpen
  \bibfield  {author} {\bibinfo {author} {\bibfnamefont {M.}~\bibnamefont
  {Balzer}}, \bibinfo {author} {\bibfnamefont {B.}~\bibnamefont {Kyung}},
  \bibinfo {author} {\bibfnamefont {D.}~\bibnamefont {S{\'{e}}n{\'{e}}chal}},
  \bibinfo {author} {\bibfnamefont {A.-M.~S.}\ \bibnamefont {Tremblay}},\ and\
  \bibinfo {author} {\bibfnamefont {M.}~\bibnamefont {Potthoff}},\ }\bibfield
  {title} {\bibinfo {title} {First-order mott transition at zero temperature in
  two dimensions: Variational plaquette study},\ }\href
  {https://doi.org/10.1209/0295-5075/85/17002} {\bibfield  {journal} {\bibinfo
  {journal} {{EPL} (Europhysics Letters)}\ }\textbf {\bibinfo {volume} {85}},\
  \bibinfo {pages} {17002} (\bibinfo {year} {2009})}\BibitemShut {NoStop}%
\bibitem [{\citenamefont {Seki}\ \emph {et~al.}(2018)\citenamefont {Seki},
  \citenamefont {Shirakawa},\ and\ \citenamefont {Yunoki}}]{Seki2018}%
  \BibitemOpen
  \bibfield  {author} {\bibinfo {author} {\bibfnamefont {K.}~\bibnamefont
  {Seki}}, \bibinfo {author} {\bibfnamefont {T.}~\bibnamefont {Shirakawa}},\
  and\ \bibinfo {author} {\bibfnamefont {S.}~\bibnamefont {Yunoki}},\
  }\bibfield  {title} {\bibinfo {title} {Variational cluster approach to
  thermodynamic properties of interacting fermions at finite temperatures: A
  case study of the two-dimensional single-band hubbard model at half
  filling},\ }\href {https://doi.org/10.1103/PhysRevB.98.205114} {\bibfield
  {journal} {\bibinfo  {journal} {Phys. Rev. B}\ }\textbf {\bibinfo {volume}
  {98}},\ \bibinfo {pages} {205114} (\bibinfo {year} {2018})}\BibitemShut
  {NoStop}%
\bibitem [{\citenamefont {Hohenadler}\ and\ \citenamefont
  {Assaad}(2013)}]{Hohenadler2013ROPIP}%
  \BibitemOpen
  \bibfield  {author} {\bibinfo {author} {\bibfnamefont {M.}~\bibnamefont
  {Hohenadler}}\ and\ \bibinfo {author} {\bibfnamefont {F.~F.}\ \bibnamefont
  {Assaad}},\ }\bibfield  {title} {\bibinfo {title} {Correlation effects in
  two-dimensional topological insulators},\ }\href@noop {} {\bibfield
  {journal} {\bibinfo  {journal} {Reports on Progress in Physics}\ }\textbf
  {\bibinfo {volume} {25}},\ \bibinfo {pages} {143201} (\bibinfo {year}
  {2013})}\BibitemShut {NoStop}%
\bibitem [{\citenamefont {Miyakoshi}\ and\ \citenamefont
  {Ohta}(2013)}]{miyakoshiPRB87}%
  \BibitemOpen
  \bibfield  {author} {\bibinfo {author} {\bibfnamefont {S.}~\bibnamefont
  {Miyakoshi}}\ and\ \bibinfo {author} {\bibfnamefont {Y.}~\bibnamefont
  {Ohta}},\ }\bibfield  {title} {\bibinfo {title} {Antiferromagnetic
  topological insulator state in the correlated bernevig-hughes-zhang model},\
  }\href {https://doi.org/10.1103/PhysRevB.87.195133} {\bibfield  {journal}
  {\bibinfo  {journal} {Phys. Rev. B}\ }\textbf {\bibinfo {volume} {87}},\
  \bibinfo {pages} {195133} (\bibinfo {year} {2013})}\BibitemShut {NoStop}%
\bibitem [{\citenamefont {Wang}\ \emph
  {et~al.}(2012{\natexlab{a}})\citenamefont {Wang}, \citenamefont {Dai},\ and\
  \citenamefont {Xie}}]{Wang2012EEL}%
  \BibitemOpen
  \bibfield  {author} {\bibinfo {author} {\bibfnamefont {L.}~\bibnamefont
  {Wang}}, \bibinfo {author} {\bibfnamefont {X.}~\bibnamefont {Dai}},\ and\
  \bibinfo {author} {\bibfnamefont {X.~C.}\ \bibnamefont {Xie}},\ }\bibfield
  {title} {\bibinfo {title} {{Interaction-induced topological phase transition
  in the Bernevig-Hughes-Zhang model}},\ }\href
  {http://stacks.iop.org/0295-5075/98/i=5/a=57001} {\bibfield  {journal}
  {\bibinfo  {journal} {EPL (Europhysics Letters)}\ }\textbf {\bibinfo {volume}
  {98}},\ \bibinfo {pages} {57001} (\bibinfo {year}
  {2012}{\natexlab{a}})}\BibitemShut {NoStop}%
\bibitem [{\citenamefont {Yoshida}\ \emph {et~al.}(2012)\citenamefont
  {Yoshida}, \citenamefont {Fujimoto},\ and\ \citenamefont
  {Kawakami}}]{Yoshida2012PRB}%
  \BibitemOpen
  \bibfield  {author} {\bibinfo {author} {\bibfnamefont {T.}~\bibnamefont
  {Yoshida}}, \bibinfo {author} {\bibfnamefont {S.}~\bibnamefont {Fujimoto}},\
  and\ \bibinfo {author} {\bibfnamefont {N.}~\bibnamefont {Kawakami}},\
  }\bibfield  {title} {\bibinfo {title} {{Correlation effects on a topological
  insulator at finite temperatures}},\ }\href
  {https://doi.org/10.1103/PhysRevB.85.125113} {\bibfield  {journal} {\bibinfo
  {journal} {Phys. Rev. B}\ }\textbf {\bibinfo {volume} {85}},\ \bibinfo
  {pages} {125113} (\bibinfo {year} {2012})}\BibitemShut {NoStop}%
\bibitem [{\citenamefont {Hohenadler}\ \emph {et~al.}(2011)\citenamefont
  {Hohenadler}, \citenamefont {Lang},\ and\ \citenamefont
  {Assaad}}]{Hohenadler2011}%
  \BibitemOpen
  \bibfield  {author} {\bibinfo {author} {\bibfnamefont {M.}~\bibnamefont
  {Hohenadler}}, \bibinfo {author} {\bibfnamefont {T.~C.}\ \bibnamefont
  {Lang}},\ and\ \bibinfo {author} {\bibfnamefont {F.~F.}\ \bibnamefont
  {Assaad}},\ }\bibfield  {title} {\bibinfo {title} {Correlation effects in
  quantum spin-hall insulators: A quantum monte carlo study},\ }\href
  {https://doi.org/10.1103/PhysRevLett.106.100403} {\bibfield  {journal}
  {\bibinfo  {journal} {Phys. Rev. Lett.}\ }\textbf {\bibinfo {volume} {106}},\
  \bibinfo {pages} {100403} (\bibinfo {year} {2011})}\BibitemShut {NoStop}%
\bibitem [{\citenamefont {Amaricci}\ \emph {et~al.}(2017)\citenamefont
  {Amaricci}, \citenamefont {Privitera}, \citenamefont {Petocchi},
  \citenamefont {Capone}, \citenamefont {Sangiovanni},\ and\ \citenamefont
  {Trauzettel}}]{Amaricci2017PRB}%
  \BibitemOpen
  \bibfield  {author} {\bibinfo {author} {\bibfnamefont {A.}~\bibnamefont
  {Amaricci}}, \bibinfo {author} {\bibfnamefont {L.}~\bibnamefont {Privitera}},
  \bibinfo {author} {\bibfnamefont {F.}~\bibnamefont {Petocchi}}, \bibinfo
  {author} {\bibfnamefont {M.}~\bibnamefont {Capone}}, \bibinfo {author}
  {\bibfnamefont {G.}~\bibnamefont {Sangiovanni}},\ and\ \bibinfo {author}
  {\bibfnamefont {B.}~\bibnamefont {Trauzettel}},\ }\bibfield  {title}
  {\bibinfo {title} {Edge state reconstruction from strong correlations in
  quantum spin hall insulators},\ }\href
  {https://doi.org/10.1103/PhysRevB.95.205120} {\bibfield  {journal} {\bibinfo
  {journal} {Phys. Rev. B}\ }\textbf {\bibinfo {volume} {95}},\ \bibinfo
  {pages} {205120} (\bibinfo {year} {2017})}\BibitemShut {NoStop}%
\bibitem [{\citenamefont {Budich}\ \emph {et~al.}(2012)\citenamefont {Budich},
  \citenamefont {Thomale}, \citenamefont {Li}, \citenamefont {Laubach},\ and\
  \citenamefont {Zhang}}]{Budich2012PRB}%
  \BibitemOpen
  \bibfield  {author} {\bibinfo {author} {\bibfnamefont {J.~C.}\ \bibnamefont
  {Budich}}, \bibinfo {author} {\bibfnamefont {R.}~\bibnamefont {Thomale}},
  \bibinfo {author} {\bibfnamefont {G.}~\bibnamefont {Li}}, \bibinfo {author}
  {\bibfnamefont {M.}~\bibnamefont {Laubach}},\ and\ \bibinfo {author}
  {\bibfnamefont {S.-C.}\ \bibnamefont {Zhang}},\ }\bibfield  {title} {\bibinfo
  {title} {{Fluctuation-induced topological quantum phase transitions in
  quantum spin-Hall and anomalous-Hall insulators}},\ }\href
  {https://doi.org/10.1103/PhysRevB.86.201407} {\bibfield  {journal} {\bibinfo
  {journal} {Phys. Rev. B}\ }\textbf {\bibinfo {volume} {86}},\ \bibinfo
  {pages} {201407} (\bibinfo {year} {2012})}\BibitemShut {NoStop}%
\bibitem [{\citenamefont {Amaricci}\ \emph {et~al.}(2018)\citenamefont
  {Amaricci}, \citenamefont {Valli}, \citenamefont {Sangiovanni}, \citenamefont
  {Trauzettel},\ and\ \citenamefont {Capone}}]{Amaricci2018PRB}%
  \BibitemOpen
  \bibfield  {author} {\bibinfo {author} {\bibfnamefont {A.}~\bibnamefont
  {Amaricci}}, \bibinfo {author} {\bibfnamefont {A.}~\bibnamefont {Valli}},
  \bibinfo {author} {\bibfnamefont {G.}~\bibnamefont {Sangiovanni}}, \bibinfo
  {author} {\bibfnamefont {B.}~\bibnamefont {Trauzettel}},\ and\ \bibinfo
  {author} {\bibfnamefont {M.}~\bibnamefont {Capone}},\ }\bibfield  {title}
  {\bibinfo {title} {Coexistence of metallic edge states and antiferromagnetic
  ordering in correlated topological insulators},\ }\href
  {https://doi.org/10.1103/PhysRevB.98.045133} {\bibfield  {journal} {\bibinfo
  {journal} {Phys. Rev. B}\ }\textbf {\bibinfo {volume} {98}},\ \bibinfo
  {pages} {045133} (\bibinfo {year} {2018})}\BibitemShut {NoStop}%
\bibitem [{\citenamefont {Rachel}(2018)}]{Rachel2018ROPIP}%
  \BibitemOpen
  \bibfield  {author} {\bibinfo {author} {\bibfnamefont {S.}~\bibnamefont
  {Rachel}},\ }\bibfield  {title} {\bibinfo {title} {Interacting topological
  insulators: a review},\ }\href {https://doi.org/10.1088/1361-6633/aad6a6}
  {\bibfield  {journal} {\bibinfo  {journal} {Reports on Progress in Physics}\
  }\textbf {\bibinfo {volume} {81}},\ \bibinfo {pages} {116501} (\bibinfo
  {year} {2018})}\BibitemShut {NoStop}%
\bibitem [{\citenamefont {Kanamori}(1963)}]{Kanamori1963}%
  \BibitemOpen
  \bibfield  {author} {\bibinfo {author} {\bibfnamefont {J.}~\bibnamefont
  {Kanamori}},\ }\bibfield  {title} {\bibinfo {title} {{Electron Correlation
  and Ferromagnetism of Transition Metals}},\ }\href
  {https://doi.org/10.1143/PTP.30.275} {\bibfield  {journal} {\bibinfo
  {journal} {Progress of Theoretical Physics}\ }\textbf {\bibinfo {volume}
  {30}},\ \bibinfo {pages} {275} (\bibinfo {year} {1963})}\BibitemShut
  {NoStop}%
\bibitem [{\citenamefont {Georges}\ \emph {et~al.}(2013)\citenamefont
  {Georges}, \citenamefont {de' Medici},\ and\ \citenamefont
  {Mravlje}}]{Georges2013ACMP}%
  \BibitemOpen
  \bibfield  {author} {\bibinfo {author} {\bibfnamefont {A.}~\bibnamefont
  {Georges}}, \bibinfo {author} {\bibfnamefont {L.}~\bibnamefont {de'
  Medici}},\ and\ \bibinfo {author} {\bibfnamefont {J.}~\bibnamefont
  {Mravlje}},\ }\bibfield  {title} {\bibinfo {title} {{ Strong Correlations
  from Hund's Coupling.}},\ }\href@noop {} {\bibfield  {journal} {\bibinfo
  {journal} {Annu.~Rev. Condens. Matter Phys.}\ }\textbf {\bibinfo {volume}
  {45}},\ \bibinfo {pages} {137} (\bibinfo {year} {2013})}\BibitemShut
  {NoStop}%
\bibitem [{\citenamefont {de' Medici}(2011)}]{Medici2011PRB}%
  \BibitemOpen
  \bibfield  {author} {\bibinfo {author} {\bibfnamefont {L.}~\bibnamefont {de'
  Medici}},\ }\bibfield  {title} {\bibinfo {title} {Hund's coupling and its key
  role in tuning multiorbital correlations},\ }\href
  {https://doi.org/10.1103/PhysRevB.83.205112} {\bibfield  {journal} {\bibinfo
  {journal} {Phys. Rev. B}\ }\textbf {\bibinfo {volume} {83}},\ \bibinfo
  {pages} {205112} (\bibinfo {year} {2011})}\BibitemShut {NoStop}%
\bibitem [{\citenamefont {Metzner}\ and\ \citenamefont
  {Vollhardt}(1989)}]{metzvol}%
  \BibitemOpen
  \bibfield  {author} {\bibinfo {author} {\bibfnamefont {W.}~\bibnamefont
  {Metzner}}\ and\ \bibinfo {author} {\bibfnamefont {D.}~\bibnamefont
  {Vollhardt}},\ }\bibfield  {title} {\bibinfo {title} {{Correlated Lattice
  Fermions in $d=\infty{}$ Dimensions}},\ }\href
  {https://doi.org/10.1103/PhysRevLett.62.324} {\bibfield  {journal} {\bibinfo
  {journal} {Phys. Rev. Lett.}\ }\textbf {\bibinfo {volume} {62}},\ \bibinfo
  {pages} {324} (\bibinfo {year} {1989})}\BibitemShut {NoStop}%
\bibitem [{\citenamefont {Müller-Hartmann}(1989)}]{MullerHartmann1989}%
  \BibitemOpen
  \bibfield  {author} {\bibinfo {author} {\bibfnamefont {E.}~\bibnamefont
  {Müller-Hartmann}},\ }\bibfield  {title} {\bibinfo {title} {Correlated
  fermions on a lattice in high dimensions},\ }\href
  {https://doi.org/10.1007/BF01311397} {\bibfield  {journal} {\bibinfo
  {journal} {Zeitschrift für Physik B Condensed Matter}\ }\textbf {\bibinfo
  {volume} {74}},\ \bibinfo {pages} {507} (\bibinfo {year} {1989})}\BibitemShut
  {NoStop}%
\bibitem [{\citenamefont {Amaricci}\ \emph {et~al.}(2015)\citenamefont
  {Amaricci}, \citenamefont {Budich}, \citenamefont {Capone}, \citenamefont
  {Trauzettel},\ and\ \citenamefont {Sangiovanni}}]{Amaricci2015PRL}%
  \BibitemOpen
  \bibfield  {author} {\bibinfo {author} {\bibfnamefont {A.}~\bibnamefont
  {Amaricci}}, \bibinfo {author} {\bibfnamefont {J.~C.}\ \bibnamefont
  {Budich}}, \bibinfo {author} {\bibfnamefont {M.}~\bibnamefont {Capone}},
  \bibinfo {author} {\bibfnamefont {B.}~\bibnamefont {Trauzettel}},\ and\
  \bibinfo {author} {\bibfnamefont {G.}~\bibnamefont {Sangiovanni}},\
  }\bibfield  {title} {\bibinfo {title} {{First-Order Character and Observable
  Signatures of Topological Quantum Phase Transitions}},\ }\href
  {https://doi.org/10.1103/PhysRevLett.114.185701} {\bibfield  {journal}
  {\bibinfo  {journal} {Phys. Rev. Lett.}\ }\textbf {\bibinfo {volume} {114}},\
  \bibinfo {pages} {185701} (\bibinfo {year} {2015})}\BibitemShut {NoStop}%
\bibitem [{\citenamefont {Amaricci}\ \emph {et~al.}(2016)\citenamefont
  {Amaricci}, \citenamefont {Budich}, \citenamefont {Capone}, \citenamefont
  {Trauzettel},\ and\ \citenamefont {Sangiovanni}}]{Amaricci2016PRB}%
  \BibitemOpen
  \bibfield  {author} {\bibinfo {author} {\bibfnamefont {A.}~\bibnamefont
  {Amaricci}}, \bibinfo {author} {\bibfnamefont {J.~C.}\ \bibnamefont
  {Budich}}, \bibinfo {author} {\bibfnamefont {M.}~\bibnamefont {Capone}},
  \bibinfo {author} {\bibfnamefont {B.}~\bibnamefont {Trauzettel}},\ and\
  \bibinfo {author} {\bibfnamefont {G.}~\bibnamefont {Sangiovanni}},\
  }\bibfield  {title} {\bibinfo {title} {Strong correlation effects on
  topological quantum phase transitions in three dimensions},\ }\href
  {https://doi.org/10.1103/PhysRevB.93.235112} {\bibfield  {journal} {\bibinfo
  {journal} {Phys. Rev. B}\ }\textbf {\bibinfo {volume} {93}},\ \bibinfo
  {pages} {235112} (\bibinfo {year} {2016})}\BibitemShut {NoStop}%
\bibitem [{\citenamefont {Crippa}\ \emph {et~al.}(2020)\citenamefont {Crippa},
  \citenamefont {Amaricci}, \citenamefont {Wagner}, \citenamefont
  {Sangiovanni}, \citenamefont {Budich},\ and\ \citenamefont
  {Capone}}]{Weyl2020PRR}%
  \BibitemOpen
  \bibfield  {author} {\bibinfo {author} {\bibfnamefont {L.}~\bibnamefont
  {Crippa}}, \bibinfo {author} {\bibfnamefont {A.}~\bibnamefont {Amaricci}},
  \bibinfo {author} {\bibfnamefont {N.}~\bibnamefont {Wagner}}, \bibinfo
  {author} {\bibfnamefont {G.}~\bibnamefont {Sangiovanni}}, \bibinfo {author}
  {\bibfnamefont {J.~C.}\ \bibnamefont {Budich}},\ and\ \bibinfo {author}
  {\bibfnamefont {M.}~\bibnamefont {Capone}},\ }\bibfield  {title} {\bibinfo
  {title} {Nonlocal annihilation of weyl fermions in correlated systems},\
  }\href {https://doi.org/10.1103/PhysRevResearch.2.012023} {\bibfield
  {journal} {\bibinfo  {journal} {Phys. Rev. Research}\ }\textbf {\bibinfo
  {volume} {2}},\ \bibinfo {pages} {012023} (\bibinfo {year}
  {2020})}\BibitemShut {NoStop}%
\bibitem [{\citenamefont {Maier}\ \emph {et~al.}(2005)\citenamefont {Maier},
  \citenamefont {Jarrell}, \citenamefont {Pruschke},\ and\ \citenamefont
  {Hettler}}]{Maier2005}%
  \BibitemOpen
  \bibfield  {author} {\bibinfo {author} {\bibfnamefont {T.}~\bibnamefont
  {Maier}}, \bibinfo {author} {\bibfnamefont {M.}~\bibnamefont {Jarrell}},
  \bibinfo {author} {\bibfnamefont {T.}~\bibnamefont {Pruschke}},\ and\
  \bibinfo {author} {\bibfnamefont {M.~H.}\ \bibnamefont {Hettler}},\
  }\bibfield  {title} {\bibinfo {title} {Quantum cluster theories},\ }\href
  {https://doi.org/10.1103/RevModPhys.77.1027} {\bibfield  {journal} {\bibinfo
  {journal} {Rev. Mod. Phys.}\ }\textbf {\bibinfo {volume} {77}},\ \bibinfo
  {pages} {1027} (\bibinfo {year} {2005})}\BibitemShut {NoStop}%
\bibitem [{\citenamefont {Kotliar}\ \emph {et~al.}(2001)\citenamefont
  {Kotliar}, \citenamefont {Savrasov}, \citenamefont {P\'alsson},\ and\
  \citenamefont {Biroli}}]{Kotliar2001}%
  \BibitemOpen
  \bibfield  {author} {\bibinfo {author} {\bibfnamefont {G.}~\bibnamefont
  {Kotliar}}, \bibinfo {author} {\bibfnamefont {S.~Y.}\ \bibnamefont
  {Savrasov}}, \bibinfo {author} {\bibfnamefont {G.}~\bibnamefont
  {P\'alsson}},\ and\ \bibinfo {author} {\bibfnamefont {G.}~\bibnamefont
  {Biroli}},\ }\bibfield  {title} {\bibinfo {title} {Cellular dynamical mean
  field approach to strongly correlated systems},\ }\href
  {https://doi.org/10.1103/PhysRevLett.87.186401} {\bibfield  {journal}
  {\bibinfo  {journal} {Phys. Rev. Lett.}\ }\textbf {\bibinfo {volume} {87}},\
  \bibinfo {pages} {186401} (\bibinfo {year} {2001})}\BibitemShut {NoStop}%
\bibitem [{\citenamefont {Biroli}\ and\ \citenamefont
  {Kotliar}(2002)}]{Biroli2002}%
  \BibitemOpen
  \bibfield  {author} {\bibinfo {author} {\bibfnamefont {G.}~\bibnamefont
  {Biroli}}\ and\ \bibinfo {author} {\bibfnamefont {G.}~\bibnamefont
  {Kotliar}},\ }\bibfield  {title} {\bibinfo {title} {Cluster methods for
  strongly correlated electron systems},\ }\href
  {https://doi.org/10.1103/PhysRevB.65.155112} {\bibfield  {journal} {\bibinfo
  {journal} {Phys. Rev. B}\ }\textbf {\bibinfo {volume} {65}},\ \bibinfo
  {pages} {155112} (\bibinfo {year} {2002})}\BibitemShut {NoStop}%
\bibitem [{\citenamefont {Potthoff}(2003)}]{Potthoff2003}%
  \BibitemOpen
  \bibfield  {author} {\bibinfo {author} {\bibfnamefont {M.}~\bibnamefont
  {Potthoff}},\ }\bibfield  {title} {\bibinfo {title} {Self-energy-functional
  approach to systems of correlated electrons},\ }\href
  {https://doi.org/10.1140/epjb/e2003-00121-8} {\bibfield  {journal} {\bibinfo
  {journal} {The European Physical Journal B - Condensed Matter and Complex
  Systems}\ }\textbf {\bibinfo {volume} {32}},\ \bibinfo {pages} {429}
  (\bibinfo {year} {2003})}\BibitemShut {NoStop}%
\bibitem [{\citenamefont {Amaricci}\ \emph {et~al.}(2021)\citenamefont
  {Amaricci}, \citenamefont {Crippa}, \citenamefont {Scazzola}, \citenamefont
  {Petocchi}, \citenamefont {Mazza}, \citenamefont {de~Medici},\ and\
  \citenamefont {Capone}}]{amaricci2021edipack}%
  \BibitemOpen
  \bibfield  {author} {\bibinfo {author} {\bibfnamefont {A.}~\bibnamefont
  {Amaricci}}, \bibinfo {author} {\bibfnamefont {L.}~\bibnamefont {Crippa}},
  \bibinfo {author} {\bibfnamefont {A.}~\bibnamefont {Scazzola}}, \bibinfo
  {author} {\bibfnamefont {F.}~\bibnamefont {Petocchi}}, \bibinfo {author}
  {\bibfnamefont {G.}~\bibnamefont {Mazza}}, \bibinfo {author} {\bibfnamefont
  {L.}~\bibnamefont {de~Medici}},\ and\ \bibinfo {author} {\bibfnamefont
  {M.}~\bibnamefont {Capone}},\ }\href@noop {} {\bibinfo {title} {Edipack: A
  parallel exact diagonalization package for quantum impurity problems}}
  (\bibinfo {year} {2021}),\ \Eprint {https://arxiv.org/abs/2105.06806}
  {arXiv:2105.06806 [physics.comp-ph]} \BibitemShut {NoStop}%
\bibitem [{\citenamefont {S\'en\'echal}(2010)}]{Senechal2010}%
  \BibitemOpen
  \bibfield  {author} {\bibinfo {author} {\bibfnamefont {D.}~\bibnamefont
  {S\'en\'echal}},\ }\bibfield  {title} {\bibinfo {title} {Bath optimization in
  the cellular dynamical mean-field theory},\ }\href
  {https://doi.org/10.1103/PhysRevB.81.235125} {\bibfield  {journal} {\bibinfo
  {journal} {Phys. Rev. B}\ }\textbf {\bibinfo {volume} {81}},\ \bibinfo
  {pages} {235125} (\bibinfo {year} {2010})}\BibitemShut {NoStop}%
\bibitem [{\citenamefont {{Civelli}}(2006)}]{CivelliThesis}%
  \BibitemOpen
  \bibfield  {author} {\bibinfo {author} {\bibfnamefont {M.}~\bibnamefont
  {{Civelli}}},\ }\emph {\bibinfo {title} {Investigation of strongly correlated
  electron systems with cellular dynamical mean field theory}},\ \href
  {https://arxiv.org/pdf/0710.2802.pdf} {Ph.D. thesis},\ \bibinfo  {school}
  {Rutgers The State University of New Jersey - New Brunswick} (\bibinfo {year}
  {2006})\BibitemShut {NoStop}%
\bibitem [{\citenamefont {Koch}\ \emph {et~al.}(2008)\citenamefont {Koch},
  \citenamefont {Sangiovanni},\ and\ \citenamefont {Gunnarsson}}]{Koch2008}%
  \BibitemOpen
  \bibfield  {author} {\bibinfo {author} {\bibfnamefont {E.}~\bibnamefont
  {Koch}}, \bibinfo {author} {\bibfnamefont {G.}~\bibnamefont {Sangiovanni}},\
  and\ \bibinfo {author} {\bibfnamefont {O.}~\bibnamefont {Gunnarsson}},\
  }\bibfield  {title} {\bibinfo {title} {Sum rules and bath parametrization for
  quantum cluster theories},\ }\href
  {https://doi.org/10.1103/PhysRevB.78.115102} {\bibfield  {journal} {\bibinfo
  {journal} {Phys. Rev. B}\ }\textbf {\bibinfo {volume} {78}},\ \bibinfo
  {pages} {115102} (\bibinfo {year} {2008})}\BibitemShut {NoStop}%
\bibitem [{\citenamefont {{Crippa}}(2020)}]{CrippaThesis}%
  \BibitemOpen
  \bibfield  {author} {\bibinfo {author} {\bibfnamefont {L.}~\bibnamefont
  {{Crippa}}},\ }\emph {\bibinfo {title} {Local and non-local correlations in
  Topological Insulators and Weyl Semimetals}},\ \href
  {http://hdl.handle.net/20.500.11767/114413} {Ph.D. thesis},\ \bibinfo
  {school} {SISSA} (\bibinfo {year} {2020})\BibitemShut {NoStop}%
\bibitem [{\citenamefont {S\'en\'echal}\ \emph {et~al.}(2000)\citenamefont
  {S\'en\'echal}, \citenamefont {Perez},\ and\ \citenamefont
  {Pioro-Ladri\`ere}}]{Senechal2000}%
  \BibitemOpen
  \bibfield  {author} {\bibinfo {author} {\bibfnamefont {D.}~\bibnamefont
  {S\'en\'echal}}, \bibinfo {author} {\bibfnamefont {D.}~\bibnamefont
  {Perez}},\ and\ \bibinfo {author} {\bibfnamefont {M.}~\bibnamefont
  {Pioro-Ladri\`ere}},\ }\bibfield  {title} {\bibinfo {title} {Spectral weight
  of the hubbard model through cluster perturbation theory},\ }\href
  {https://doi.org/10.1103/PhysRevLett.84.522} {\bibfield  {journal} {\bibinfo
  {journal} {Phys. Rev. Lett.}\ }\textbf {\bibinfo {volume} {84}},\ \bibinfo
  {pages} {522} (\bibinfo {year} {2000})}\BibitemShut {NoStop}%
\bibitem [{\citenamefont {Stanescu}\ and\ \citenamefont
  {Kotliar}(2006)}]{Stanescu2006}%
  \BibitemOpen
  \bibfield  {author} {\bibinfo {author} {\bibfnamefont {T.~D.}\ \bibnamefont
  {Stanescu}}\ and\ \bibinfo {author} {\bibfnamefont {G.}~\bibnamefont
  {Kotliar}},\ }\bibfield  {title} {\bibinfo {title} {Fermi arcs and hidden
  zeros of the green function in the pseudogap state},\ }\href
  {https://doi.org/10.1103/PhysRevB.74.125110} {\bibfield  {journal} {\bibinfo
  {journal} {Phys. Rev. B}\ }\textbf {\bibinfo {volume} {74}},\ \bibinfo
  {pages} {125110} (\bibinfo {year} {2006})}\BibitemShut {NoStop}%
\bibitem [{\citenamefont {Zheng}\ and\ \citenamefont
  {Hofstetter}(2018)}]{Hofstetter2018}%
  \BibitemOpen
  \bibfield  {author} {\bibinfo {author} {\bibfnamefont {J.-H.}\ \bibnamefont
  {Zheng}}\ and\ \bibinfo {author} {\bibfnamefont {W.}~\bibnamefont
  {Hofstetter}},\ }\bibfield  {title} {\bibinfo {title} {Topological invariant
  for two-dimensional open systems},\ }\href
  {https://doi.org/10.1103/PhysRevB.97.195434} {\bibfield  {journal} {\bibinfo
  {journal} {Phys. Rev. B}\ }\textbf {\bibinfo {volume} {97}},\ \bibinfo
  {pages} {195434} (\bibinfo {year} {2018})}\BibitemShut {NoStop}%
\bibitem [{\citenamefont {Markov}\ \emph {et~al.}(2019)\citenamefont {Markov},
  \citenamefont {Rohringer},\ and\ \citenamefont {Rubtsov}}]{Markov2019}%
  \BibitemOpen
  \bibfield  {author} {\bibinfo {author} {\bibfnamefont {A.~A.}\ \bibnamefont
  {Markov}}, \bibinfo {author} {\bibfnamefont {G.}~\bibnamefont {Rohringer}},\
  and\ \bibinfo {author} {\bibfnamefont {A.~N.}\ \bibnamefont {Rubtsov}},\
  }\bibfield  {title} {\bibinfo {title} {Robustness of the topological
  quantization of the hall conductivity for correlated lattice electrons at
  finite temperatures},\ }\href {https://doi.org/10.1103/PhysRevB.100.115102}
  {\bibfield  {journal} {\bibinfo  {journal} {Phys. Rev. B}\ }\textbf {\bibinfo
  {volume} {100}},\ \bibinfo {pages} {115102} (\bibinfo {year}
  {2019})}\BibitemShut {NoStop}%
\bibitem [{\citenamefont {Thunström}\ and\ \citenamefont
  {Held}(2019)}]{thunstrom2019topology}%
  \BibitemOpen
  \bibfield  {author} {\bibinfo {author} {\bibfnamefont {P.}~\bibnamefont
  {Thunström}}\ and\ \bibinfo {author} {\bibfnamefont {K.}~\bibnamefont
  {Held}},\ }\href@noop {} {\bibinfo {title} {Topology of smb$_6$ determined by
  dynamical mean field theory}} (\bibinfo {year} {2019}),\ \Eprint
  {https://arxiv.org/abs/1907.03899} {arXiv:1907.03899 [cond-mat.str-el]}
  \BibitemShut {NoStop}%
\bibitem [{\citenamefont {Wang}\ and\ \citenamefont
  {Zhang}(2012)}]{WangZhang2012}%
  \BibitemOpen
  \bibfield  {author} {\bibinfo {author} {\bibfnamefont {Z.}~\bibnamefont
  {Wang}}\ and\ \bibinfo {author} {\bibfnamefont {S.-C.}\ \bibnamefont
  {Zhang}},\ }\bibfield  {title} {\bibinfo {title} {Simplified topological
  invariants for interacting insulators},\ }\href
  {https://doi.org/10.1103/PhysRevX.2.031008} {\bibfield  {journal} {\bibinfo
  {journal} {Phys. Rev. X}\ }\textbf {\bibinfo {volume} {2}},\ \bibinfo {pages}
  {031008} (\bibinfo {year} {2012})}\BibitemShut {NoStop}%
\bibitem [{\citenamefont {Wang}\ \emph
  {et~al.}(2012{\natexlab{b}})\citenamefont {Wang}, \citenamefont {Qi},\ and\
  \citenamefont {Zhang}}]{Wang2012}%
  \BibitemOpen
  \bibfield  {author} {\bibinfo {author} {\bibfnamefont {Z.}~\bibnamefont
  {Wang}}, \bibinfo {author} {\bibfnamefont {X.-L.}\ \bibnamefont {Qi}},\ and\
  \bibinfo {author} {\bibfnamefont {S.-C.}\ \bibnamefont {Zhang}},\ }\bibfield
  {title} {\bibinfo {title} {Topological invariants for interacting topological
  insulators with inversion symmetry},\ }\href
  {https://doi.org/10.1103/PhysRevB.85.165126} {\bibfield  {journal} {\bibinfo
  {journal} {Phys. Rev. B}\ }\textbf {\bibinfo {volume} {85}},\ \bibinfo
  {pages} {165126} (\bibinfo {year} {2012}{\natexlab{b}})}\BibitemShut
  {NoStop}%
\bibitem [{Note1()}]{Note1}%
  \BibitemOpen
  \bibinfo {note} {The expression is manifestly non-symmetric due to the choice
  of the cluster which is aligned along the x direction, hence non-local
  self-energies enter only in the $k_x$ direction. For a more symmetric
  cluster, the $k_x$ and $k_y$ direction would display the same
  renormalization.}\BibitemShut {Stop}%
\bibitem [{\citenamefont {Parragh}\ \emph {et~al.}(2013)\citenamefont
  {Parragh}, \citenamefont {Sangiovanni}, \citenamefont {Hansmann},
  \citenamefont {Hummel}, \citenamefont {Held},\ and\ \citenamefont
  {Toschi}}]{Parragh2013}%
  \BibitemOpen
  \bibfield  {author} {\bibinfo {author} {\bibfnamefont {N.}~\bibnamefont
  {Parragh}}, \bibinfo {author} {\bibfnamefont {G.}~\bibnamefont
  {Sangiovanni}}, \bibinfo {author} {\bibfnamefont {P.}~\bibnamefont
  {Hansmann}}, \bibinfo {author} {\bibfnamefont {S.}~\bibnamefont {Hummel}},
  \bibinfo {author} {\bibfnamefont {K.}~\bibnamefont {Held}},\ and\ \bibinfo
  {author} {\bibfnamefont {A.}~\bibnamefont {Toschi}},\ }\bibfield  {title}
  {\bibinfo {title} {Effective crystal field and fermi surface topology: A
  comparison of $d$- and $dp$-orbital models},\ }\href
  {https://doi.org/10.1103/PhysRevB.88.195116} {\bibfield  {journal} {\bibinfo
  {journal} {Phys. Rev. B}\ }\textbf {\bibinfo {volume} {88}},\ \bibinfo
  {pages} {195116} (\bibinfo {year} {2013})}\BibitemShut {NoStop}%
\bibitem [{\citenamefont {Budich}\ \emph {et~al.}(2013)\citenamefont {Budich},
  \citenamefont {Trauzettel},\ and\ \citenamefont
  {Sangiovanni}}]{Budich2013PRB}%
  \BibitemOpen
  \bibfield  {author} {\bibinfo {author} {\bibfnamefont {J.~C.}\ \bibnamefont
  {Budich}}, \bibinfo {author} {\bibfnamefont {B.}~\bibnamefont {Trauzettel}},\
  and\ \bibinfo {author} {\bibfnamefont {G.}~\bibnamefont {Sangiovanni}},\
  }\bibfield  {title} {\bibinfo {title} {{Fluctuation-driven topological Hund
  insulators}},\ }\href {https://doi.org/10.1103/PhysRevB.87.235104} {\bibfield
   {journal} {\bibinfo  {journal} {Phys. Rev. B}\ }\textbf {\bibinfo {volume}
  {87}},\ \bibinfo {pages} {235104} (\bibinfo {year} {2013})}\BibitemShut
  {NoStop}%
\bibitem [{\citenamefont {Werner}\ and\ \citenamefont
  {Millis}(2007)}]{Werner2007b}%
  \BibitemOpen
  \bibfield  {author} {\bibinfo {author} {\bibfnamefont {P.}~\bibnamefont
  {Werner}}\ and\ \bibinfo {author} {\bibfnamefont {A.~J.}\ \bibnamefont
  {Millis}},\ }\bibfield  {title} {\bibinfo {title} {Doping-driven mott
  transition in the one-band hubbard model},\ }\href
  {https://doi.org/10.1103/PhysRevB.75.085108} {\bibfield  {journal} {\bibinfo
  {journal} {Phys. Rev. B}\ }\textbf {\bibinfo {volume} {75}},\ \bibinfo
  {pages} {085108} (\bibinfo {year} {2007})}\BibitemShut {NoStop}%
\bibitem [{\citenamefont {Rachel}\ and\ \citenamefont
  {Le~Hur}(2010)}]{Rachel2010}%
  \BibitemOpen
  \bibfield  {author} {\bibinfo {author} {\bibfnamefont {S.}~\bibnamefont
  {Rachel}}\ and\ \bibinfo {author} {\bibfnamefont {K.}~\bibnamefont
  {Le~Hur}},\ }\bibfield  {title} {\bibinfo {title} {Topological insulators and
  mott physics from the hubbard interaction},\ }\href
  {https://doi.org/10.1103/PhysRevB.82.075106} {\bibfield  {journal} {\bibinfo
  {journal} {Phys. Rev. B}\ }\textbf {\bibinfo {volume} {82}},\ \bibinfo
  {pages} {075106} (\bibinfo {year} {2010})}\BibitemShut {NoStop}%
\bibitem [{\citenamefont {R\"uegg}\ and\ \citenamefont
  {Fiete}(2012)}]{Ruegg2012}%
  \BibitemOpen
  \bibfield  {author} {\bibinfo {author} {\bibfnamefont {A.}~\bibnamefont
  {R\"uegg}}\ and\ \bibinfo {author} {\bibfnamefont {G.~A.}\ \bibnamefont
  {Fiete}},\ }\bibfield  {title} {\bibinfo {title} {Topological order and
  semions in a strongly correlated quantum spin hall insulator},\ }\href
  {https://doi.org/10.1103/PhysRevLett.108.046401} {\bibfield  {journal}
  {\bibinfo  {journal} {Phys. Rev. Lett.}\ }\textbf {\bibinfo {volume} {108}},\
  \bibinfo {pages} {046401} (\bibinfo {year} {2012})}\BibitemShut {NoStop}%
\bibitem [{\citenamefont {Zheng}\ \emph {et~al.}(2020)\citenamefont {Zheng},
  \citenamefont {Irsigler}, \citenamefont {Jiang}, \citenamefont {Weitenberg},\
  and\ \citenamefont {Hofstetter}}]{Zheng2020}%
  \BibitemOpen
  \bibfield  {author} {\bibinfo {author} {\bibfnamefont {J.-H.}\ \bibnamefont
  {Zheng}}, \bibinfo {author} {\bibfnamefont {B.}~\bibnamefont {Irsigler}},
  \bibinfo {author} {\bibfnamefont {L.}~\bibnamefont {Jiang}}, \bibinfo
  {author} {\bibfnamefont {C.}~\bibnamefont {Weitenberg}},\ and\ \bibinfo
  {author} {\bibfnamefont {W.}~\bibnamefont {Hofstetter}},\ }\bibfield  {title}
  {\bibinfo {title} {Measuring an interaction-induced topological phase
  transition via the single-particle density matrix},\ }\href
  {https://doi.org/10.1103/PhysRevA.101.013631} {\bibfield  {journal} {\bibinfo
   {journal} {Phys. Rev. A}\ }\textbf {\bibinfo {volume} {101}},\ \bibinfo
  {pages} {013631} (\bibinfo {year} {2020})}\BibitemShut {NoStop}%
\bibitem [{\citenamefont {Mertz}\ \emph {et~al.}(2019)\citenamefont {Mertz},
  \citenamefont {Zantout},\ and\ \citenamefont {Valent\'{\i}}}]{Mertz2019}%
  \BibitemOpen
  \bibfield  {author} {\bibinfo {author} {\bibfnamefont {T.}~\bibnamefont
  {Mertz}}, \bibinfo {author} {\bibfnamefont {K.}~\bibnamefont {Zantout}},\
  and\ \bibinfo {author} {\bibfnamefont {R.}~\bibnamefont {Valent\'{\i}}},\
  }\bibfield  {title} {\bibinfo {title} {Statistical analysis of the chern
  number in the interacting haldane-hubbard model},\ }\href
  {https://doi.org/10.1103/PhysRevB.100.125111} {\bibfield  {journal} {\bibinfo
   {journal} {Phys. Rev. B}\ }\textbf {\bibinfo {volume} {100}},\ \bibinfo
  {pages} {125111} (\bibinfo {year} {2019})}\BibitemShut {NoStop}%
\bibitem [{\citenamefont {Pizarro}\ \emph {et~al.}(2020)\citenamefont
  {Pizarro}, \citenamefont {Adler}, \citenamefont {Zantout}, \citenamefont
  {Mertz}, \citenamefont {Barone}, \citenamefont {Valentí}, \citenamefont
  {Sangiovanni},\ and\ \citenamefont {Wehling}}]{Pizarro2020}%
  \BibitemOpen
  \bibfield  {author} {\bibinfo {author} {\bibfnamefont {J.~M.}\ \bibnamefont
  {Pizarro}}, \bibinfo {author} {\bibfnamefont {S.}~\bibnamefont {Adler}},
  \bibinfo {author} {\bibfnamefont {K.}~\bibnamefont {Zantout}}, \bibinfo
  {author} {\bibfnamefont {T.}~\bibnamefont {Mertz}}, \bibinfo {author}
  {\bibfnamefont {P.}~\bibnamefont {Barone}}, \bibinfo {author} {\bibfnamefont
  {R.}~\bibnamefont {Valentí}}, \bibinfo {author} {\bibfnamefont
  {G.}~\bibnamefont {Sangiovanni}},\ and\ \bibinfo {author} {\bibfnamefont
  {T.~O.}\ \bibnamefont {Wehling}},\ }\bibfield  {title} {\bibinfo {title}
  {Deconfinement of mott localized electrons into topological and
  spin--orbit-coupled dirac fermions},\ }\href
  {https://doi.org/10.1038/s41535-020-00277-3} {\bibfield  {journal} {\bibinfo
  {journal} {npj Quantum Materials}\ }\textbf {\bibinfo {volume} {5}},\
  \bibinfo {pages} {79} (\bibinfo {year} {2020})}\BibitemShut {NoStop}%
\bibitem [{\citenamefont {Kane}\ and\ \citenamefont
  {Mele}(2005{\natexlab{c}})}]{KaneMele2005}%
  \BibitemOpen
  \bibfield  {author} {\bibinfo {author} {\bibfnamefont {C.~L.}\ \bibnamefont
  {Kane}}\ and\ \bibinfo {author} {\bibfnamefont {E.~J.}\ \bibnamefont
  {Mele}},\ }\bibfield  {title} {\bibinfo {title} {Quantum spin hall effect in
  graphene},\ }\href {https://doi.org/10.1103/PhysRevLett.95.226801} {\bibfield
   {journal} {\bibinfo  {journal} {Phys. Rev. Lett.}\ }\textbf {\bibinfo
  {volume} {95}},\ \bibinfo {pages} {226801} (\bibinfo {year}
  {2005}{\natexlab{c}})}\BibitemShut {NoStop}%
\bibitem [{\citenamefont {Haldane}(1988)}]{Haldane1988}%
  \BibitemOpen
  \bibfield  {author} {\bibinfo {author} {\bibfnamefont {F.~D.~M.}\
  \bibnamefont {Haldane}},\ }\bibfield  {title} {\bibinfo {title} {Model for a
  quantum hall effect without landau levels: Condensed-matter realization of
  the "parity anomaly"},\ }\href {https://doi.org/10.1103/PhysRevLett.61.2015}
  {\bibfield  {journal} {\bibinfo  {journal} {Phys. Rev. Lett.}\ }\textbf
  {\bibinfo {volume} {61}},\ \bibinfo {pages} {2015} (\bibinfo {year}
  {1988})}\BibitemShut {NoStop}%
\bibitem [{\citenamefont {Kane}\ and\ \citenamefont
  {Mele}(2005{\natexlab{d}})}]{KaneMele2005b}%
  \BibitemOpen
  \bibfield  {author} {\bibinfo {author} {\bibfnamefont {C.~L.}\ \bibnamefont
  {Kane}}\ and\ \bibinfo {author} {\bibfnamefont {E.~J.}\ \bibnamefont
  {Mele}},\ }\bibfield  {title} {\bibinfo {title} {${Z}_{2}$ topological order
  and the quantum spin hall effect},\ }\href
  {https://doi.org/10.1103/PhysRevLett.95.146802} {\bibfield  {journal}
  {\bibinfo  {journal} {Phys. Rev. Lett.}\ }\textbf {\bibinfo {volume} {95}},\
  \bibinfo {pages} {146802} (\bibinfo {year} {2005}{\natexlab{d}})}\BibitemShut
  {NoStop}%
\end{thebibliography}%

\end{document}